\documentclass[]{aa}
\usepackage{epsfig}
\usepackage{natbib}
\usepackage{txfonts}
\newcommand{\sub}[1]{_{\rm #1}}
\newcommand{\reference}[1]{}

\def\ApJ{{ApJ }}

\def\AandA{{A\&A }}

\def\MN{{MNRAS }}

\newcommand{\changed}{}
\newcommand{\newchanged}{}

\begin{document}

\title{Structure analysis of interstellar clouds: I. Improving the
  $\Delta$-variance method}
\titlerunning{Improving the
  $\Delta$-variance method}
\author{V.~Ossenkopf\inst{1,2,3}, M.~Krips\inst{1,4} \and{} J. Stutzki\inst{1}}

\institute{1. Physikalisches Institut der Universit\"at 
zu K\"oln, Z\"ulpicher Stra\ss{}e 77, 50937 K\"oln, Germany
\and
SRON Netherlands Institute for Space Research, P.O. Box 800, 9700 AV 
Groningen, Netherlands
\and
 Kapteyn Astronomical Institute, University of Groningen, PO box 800, 
 9700 AV Groningen, Netherlands
\and
Harvard-Smithsonian Center for Astrophysics, SMA project,
60 Garden Street, MS 78 Cambridge, MA  02138, USA}

\date{Received: December 23, 2002; accepted February 22, 2008}

\abstract
% Introduction
{
The $\Delta$-variance analysis, {\changed introduced as a wavelet-based 
measure for the statistical scaling of structures in astronomical maps}, has 
proven to be an efficient and accurate method of characterising 
the power spectrum of interstellar turbulence.
It has been applied to observed molecular cloud maps and corresponding
simulated maps generated from turbulent cloud models. 
The implementation presently in use,
however, has several shortcomings. It does not take into account
the different degree of uncertainty of map values for different
points in the map, its computation by
convolution in spatial coordinates is very time-consuming, and the
selection of the wavelet is somewhat arbitrary and does not
provide an exact value for the scales traced.
}
%Aims
{
We propose and test an improved $\Delta$-variance algorithm for two-dimensional
data sets, {\changed which is applicable to maps with variable error bars and
which can be quickly computed in Fourier space. We calibrate the
spatial resolution of the $\Delta$-variance spectra.}
}
%Methods
{
{\changed The new $\Delta$-variance algorithm is based on an appropriate 
filtering of the data} in Fourier space. It uses a supplementary significance
function by which each data point is weighted. This allows us to
distinguish the influence of variable noise from the actual small-scale
structure in the maps and it helps for dealing with the boundary problem
in non-periodic and/or irregularly bounded maps. {\changed Applying the
method to artificial maps with variable noise shows that we can extend 
the dynamic range for a reliable determination of the spectral index
considerably.
We try several wavelets and test their spatial sensitivity using artificial 
maps with well known structure sizes. Performing the convolution
in Fourier space provides a major speed-up of the analysis.}
}
%Results
{
It turns out that different wavelets show different strengths with respect
to detecting characteristic structures and spectral indices, i.e.
different aspects of map structures. 
As a reasonable universal compromise for the optimum $\Delta$-variance
filter, we propose the Mexican-hat filter with a ratio between the diameters 
of the core and the annulus of 1.5. {\changed When the main focus lies on 
measuring the spectral index, the French-hat filter with a diameter
ratio of about 2.3 is also suitable. In paper II we exploit the strength
of the new method by applying it to different astronomical data.}
}
%Conclusions
{}
\keywords{Methods: data analysis -- Methods: statistical --
ISM: clouds -- ISM: structure}

\maketitle

\section{Introduction}

The interstellar medium is highly turbulent and {\changed turbulent motions
determine} the evolution of
interstellar clouds. The turbulent pressure is partially able to
support them against gravitational collapse \citep{Klessen}; 
turbulent shocks create and dissolve dense clumps in molecular clouds
or even whole clouds \citep{Javier}, the turbulent mass transport
modifies their chemical evolution \citep{LeBourlot}, and
the irregular turbulent structure determines their penetration 
by UV radiation \citep{PDR}. Thus, the complex dynamic structure
on all scales resulting from turbulence has important implications
for many aspects of the astrophysics of the interstellar matter. 

Whereas many observations reveal the complexity of
the structure of the interstellar medium, most models of interstellar
clouds are still based on simple geometrical configurations.
A first step towards a better understanding of interstellar turbulence and
towards building more realistic models of interstellar clouds is to identify
model structures characterised by a limited set of parameters which can
be quantified by comparison with observed cloud images. As many
aspects of observed interstellar clouds can be described by
fractal properties \citep{Combes}, a promising first approach to
a parametric description is given by exponents of scaling relations.

Motivated by the similarity of observed interstellar cloud images with 
the structure of {\it fractional Brownian motion} (fBm, see Sect.
\ref{sect_pertestdata}) fractals, which are characterised by the 
single number of the exponent of the power spectrum, \citet{Stutzki}
developed the $\Delta$-variance analysis as a tool to 
measure the structural scaling behaviour of observed images. 

The $\Delta$-variance is a type of averaged wavelet transform that
measures the variance in a structure $f(\vec{r})$ on a given scale
$l$ by filtering it by a spherically symmetric down-up-down function
of size $l$ (Zielinsky \& Stutzki 1999). The $\Delta$-variance 
analysis was successfully applied to several observational data sets:
\citet{Stutzki} studied a CO map of the Outer Galaxy, \citet{Bensch}
investigated a series of nearby star-forming clouds and a number of
nested maps in different CO isotopes from the Polaris Flare, 
\citet{Huber} performed a systematic study of a large set of Galactic
CO maps, and {\changed \citet{Sun} analysed maps of the Perseus molecular cloud
taken in various tracers and including the analysis of velocity channels.
The intensity maps of most clouds resulted in}
power-law $\Delta$-variance spectra
with exponents between 0.5 and 1.3. Mac Low \& Ossenkopf 
(2000) and Ossenkopf (2002) applied the $\Delta$-variance analysis to simulations of
interstellar turbulence to compare the scaling behaviour of the simulations
with that of observed maps. It became however obvious that,
aside from the spectral index, deviations from a power law
on particular scales should be studied as well because they provide
significant information on the physical processes on these scales.
Thus the $\Delta$-variance analysis is to be optimised with
respect to its capabilities of the corresponding scale detection.

We propose in this paper a number of improvements to the $\Delta$-variance
optimising its sensitivity, its applicability to arbitrary data sets,
and the speed of its computation. The critical quantity for the
detection of pronounced scales in a structure is the shape
of the wavelet filter function. The spherically symmetric down-up-down function
introduced by \citet{Stutzki} is an obvious first choice. However, other
wavelet shapes offer attractive alternatives. 

For infinitely extended or for periodic structures the fastest way of
numerically calculating the $\Delta$-variance is given by a Fourier transform
of the image. However, observed maps typically have a finite size, often
even cutting the observed clouds at the map boundary, and Fourier-based
methods run into the well known problems of artificial structure being
introduced by these edge effects. \citet{Bensch} thus implemented the
$\Delta$-variance by a numerical treatment in the spatial domain.
Calculating a two-dimensional convolution in the spatial domain, 
however, results in a rather slow computation. An additional complication
in observed data comes from the fact that the signal-to-noise
ratio is often not uniform across the mapped area. 
{\changed A particular example of maps with strongly variable data 
reliability are line centroid velocity maps.
Here, the accuracy of the centroid velocity always depends on the line
intensity. \citet{MLO} have shown that the ``traditional´´ $\Delta$-variance
analysis may fail in this case.}
To relieve these problems and concerns, we introduce a supplementary function
into the $\Delta$-variance analysis which is used to weight the data points
in the spatial map according to their significance. This helps to derive
correct contributions of data points with a different signal-to-noise
ratio to the structure information on a particular spatial scale
and it allows us to calculate the $\Delta$-variance in Fourier space and
thus to make use of the numerical advantages of the fast Fourier transform 
algorithm.

After revising the fundamental properties of the $\Delta$-variance and
defining appropriate images to test the method in Sect. 2, we introduce
the concepts of the improved $\Delta$-variance including a weighting
function in Sect. 3 and optimise it with respect to the wavelet filter function
in Sect. 4, where we also verify its performance by extensively testing
it against the test structures. {\newchanged A combined test of the 
optimised filter function and the significance function in the case of 
noisy data is presented in the Appendix. Sect. 5} summarises our
findings providing recommendations for the optimum method and wavelet
to use. In a second paper, we test the capability of the new method
applying it to simulations of interstellar turbulence and observed molecular
line maps exploiting the improved sensitivity to derive general properties
of interstellar turbulence.

\section{The starting point}

\subsection{The $\Delta$-variance}
\label{sect_mathintro}

The $\Delta$-variance analysis was comprehensively introduced by
\citet{Stutzki} and \citet{Bensch}.  Here, we 
only repeat those equations which are essential to understand the
extensions proposed in Sects. \ref{sect_edge} and \ref{sect_filter}.
Although  the $\Delta$-variance can be used 
in principle for an arbitrary number of dimensions we restrict
ourselves to the two-dimensional case, i.e. the analysis of
maps or images. 
%They represent the most common astrophysical data sets
%due to the natural limitation of observations to the celestial 
%coordinates. 

The $\Delta$-variance measures the amount of structure on a given scale
$l$ in a map $f(\vec{r})$ by filtering the map with a spherically symmetric 
down-up-down function of size $l$ (French-hat filter) and computing
the variance of the thus filtered map. It is given by
\begin{equation}
\sigma_\Delta^2(l)= \left\langle \left( f(\vec{r}) * {\bigodot}_l(\vec{r}) \right)^2 \right\rangle_{\vec{r}}
\end{equation}
where, the average is taken over the area of the map, 
the symbol $*$ stands for a convolution, and $\bigodot_l$ describes
the French-hat function defined as
\begin{equation}
{\bigodot}_l(\vec{r})= {4 \over \pi l^2} \left\{ \begin{array}{ll}
1 &: |\vec{r}| \le l/2\\
{-1 / 8} &: l/2 < |\vec{r}| \le 3l/2\\
0 &: |\vec{r}| > 3l/2 \end{array} \right.
\label{eq_frenchhat}
\end{equation}
Thus the filter consists of a positive core and a negative annulus where
the width of the annulus agrees with the diameter of the core and
{\changed the absolute values in each of them are inversely proportional to their areas
so that they both have an integral weight of unity}
\footnote{Following the original definition by \citet{Stutzki} our
$\Delta$-variance is higher by the constant factor of $2\pi$ 
than the definition used by \citet{Bensch}.}.

In a more general picture one can consider the filter function
as a wavelet composed of a negative and a positive part both normalised
to integral values of unity so that the overall filter has a vanishing
integral. Using an arbitrary diameter ratio between the
annulus and the core $v$ we can write
\begin{eqnarray}
{\bigodot}_l(\vec{r})&=&\bigodot_{l, {\rm core}}(\vec{r})
- \bigodot_{l, {\rm ann}}(\vec{r}) \\
\bigodot_{l, {\rm core}}(\vec{r}) &=& {4 \over \pi l^2}
\left\{ \begin{array}{ll} 1 &: |\vec{r}| \le l/2\\
0 &: |\vec{r}| > l/2 \end{array} \right. \nonumber \\
\bigodot_{l, {\rm ann}}(\vec{r}) &=& {4 \over \pi l^2}
\left\{ \begin{array}{ll} {1/(v^2-1)} &: l/2 < |\vec{r}| \le v \times l/2\\
0 &: |\vec{r}| \le l/2, |\vec{r}| > v\times l/2 \end{array} \right.
\label{fhat}
\end{eqnarray}
The ``traditional'' $\Delta$-variance filter is reproduced for a diameter
ratio $v=3$. 
We come back to discussing the choice of $v$ in Sect. 4.

Because the average distance between two points in the core and the
annulus of the filter is close to the length $l$ 
(see Sect. \ref{sect_effectivelength}), the convolved map
only retains variations on that scale whereas variations on smaller
and larger scales are suppressed.  The $\Delta$-variance as the
variance of the convolved map thus measures the amount of structural
variation on the scale $l$. Plotting the $\Delta$-variance
as a function of the filter size $l$ then provides a spectrum
showing the relative amount of structure in a given map as a
function of the structure size.

The filter convolution and computation of the $\Delta$-variance
can be easily performed in Fourier space where they are
reduced to a simple multiplication and integration.
This directly relates the $\Delta$-variance
to the power spectrum. If $P(|\vec{k}|)$ is 
the radially averaged power spectrum of the structure 
$f(\vec{r})$, the $\Delta$-variance is given by
\begin{equation}
\sigma_\Delta^2(l)= 2\pi \; \int_0^\infty P(|\vec{k}|)\; 
\left| \tilde{\bigodot}_l(|\vec{k}|) \right|^2 |\vec{k}| \; d|\vec{k}|
\label{eq_deltafourier}
\end{equation}
where $\tilde{\bigodot}_l$ is the Fourier transform of the
filter function with the size $l$ and $\vec{k}$
denotes the spatial frequency or wavenumber.
If the power spectrum is given by a power law, $P(|\vec{k}|)\propto
|\vec{k}|^{-\zeta}$, the $\Delta$-variance also follows a power law
$\sigma_\Delta^2 \propto l^\alpha$ with $\alpha={\zeta-2}$ within 
the exponential range $0\le \zeta < 6$ \citep{Stutzki}\footnote{
Please, note the difference to the often used energy spectrum $P(k)$
which is obtained by angular integration of $P(\vec{k})$ and has
the spectral index $\zeta\sub{int}=\zeta-1$.}.

Thus, the $\Delta$-variance shows in principle only information
that is also contained in the power spectrum. The main advantage of the
$\Delta$-variance method compared to the direct computation of the power
spectrum results from the smooth filter shape which provides
a very robust way for an angular average, insensitivity to 
singular variations, and independence of gridding and finite map 
size effects. It provides a good separation of different
effects based on their characteristic scale, e.g. a clear distinction
between observational noise, structure blurring by the finite 
telescope beam, and the internal scaling of the astrophysical source.
A detailed computation of the influence of finite map sizes and
telescope blurring was provided by \citet{Bensch}.

{\changed However, when applying Eq. (\ref{eq_deltafourier}) to compute
the $\Delta$-variance it inherits a main drawback from} the power spectrum -- it
implicitly assumes a periodic continuation in the Fourier transform
although most astrophysical observations show no periodicity.
Sect. \ref{sect_edge}
thus deals with the implications of this assumption and possible
ways to overcome the resulting limitations.

\begin{figure*}[ht]
\epsfig{file=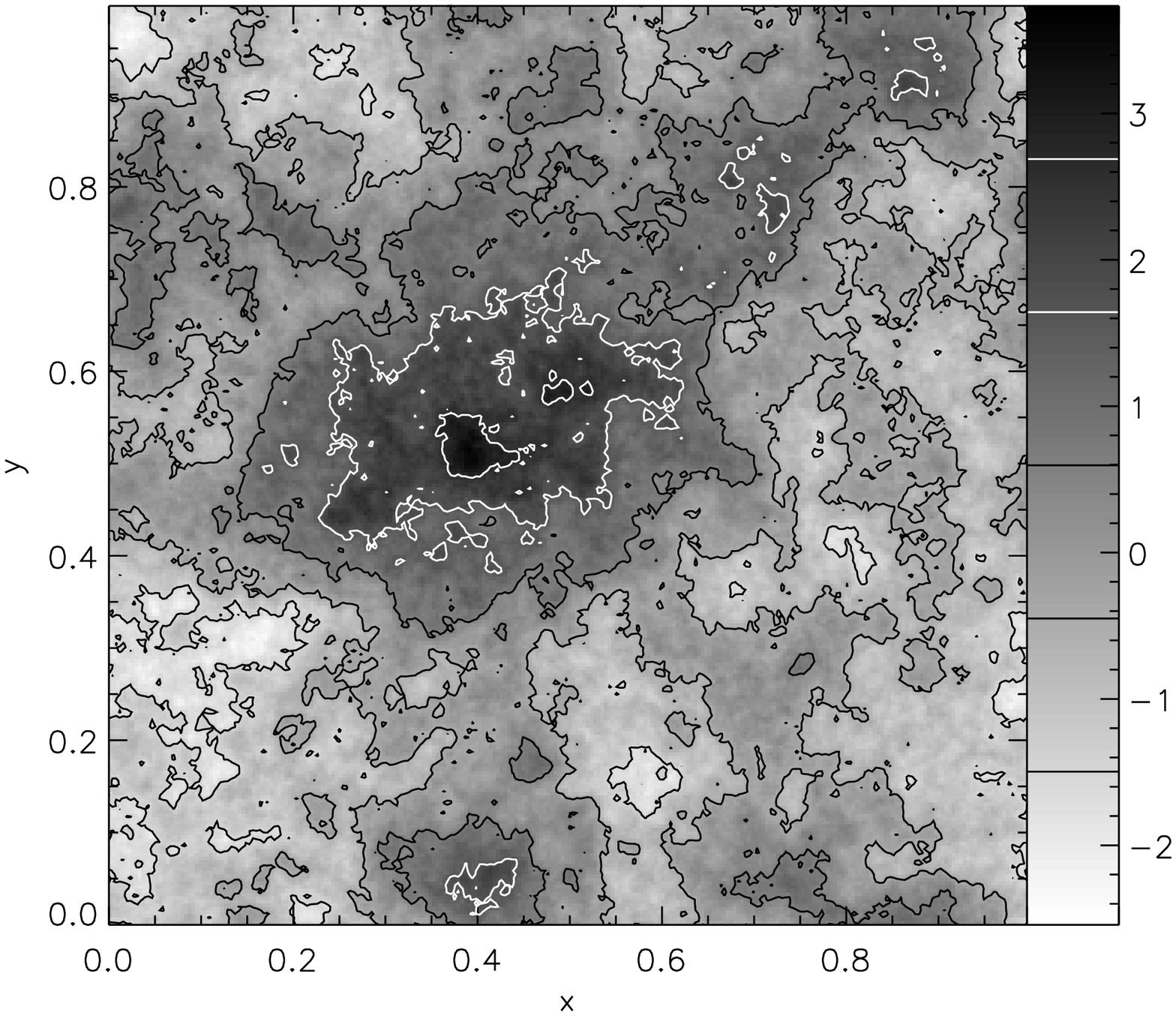, angle=90, width=6cm}
\epsfig{file=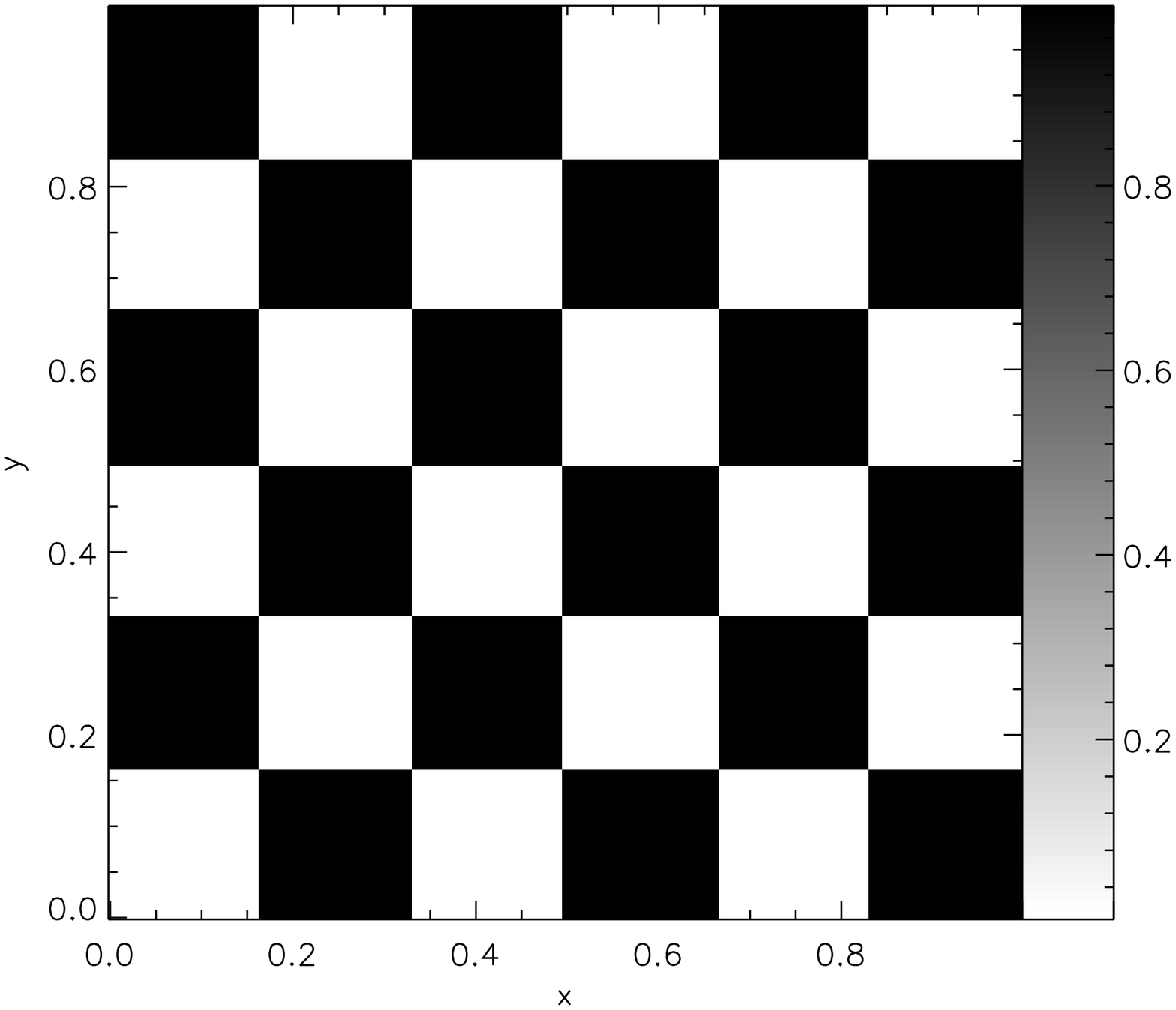, angle=90, width=6cm}
\epsfig{file=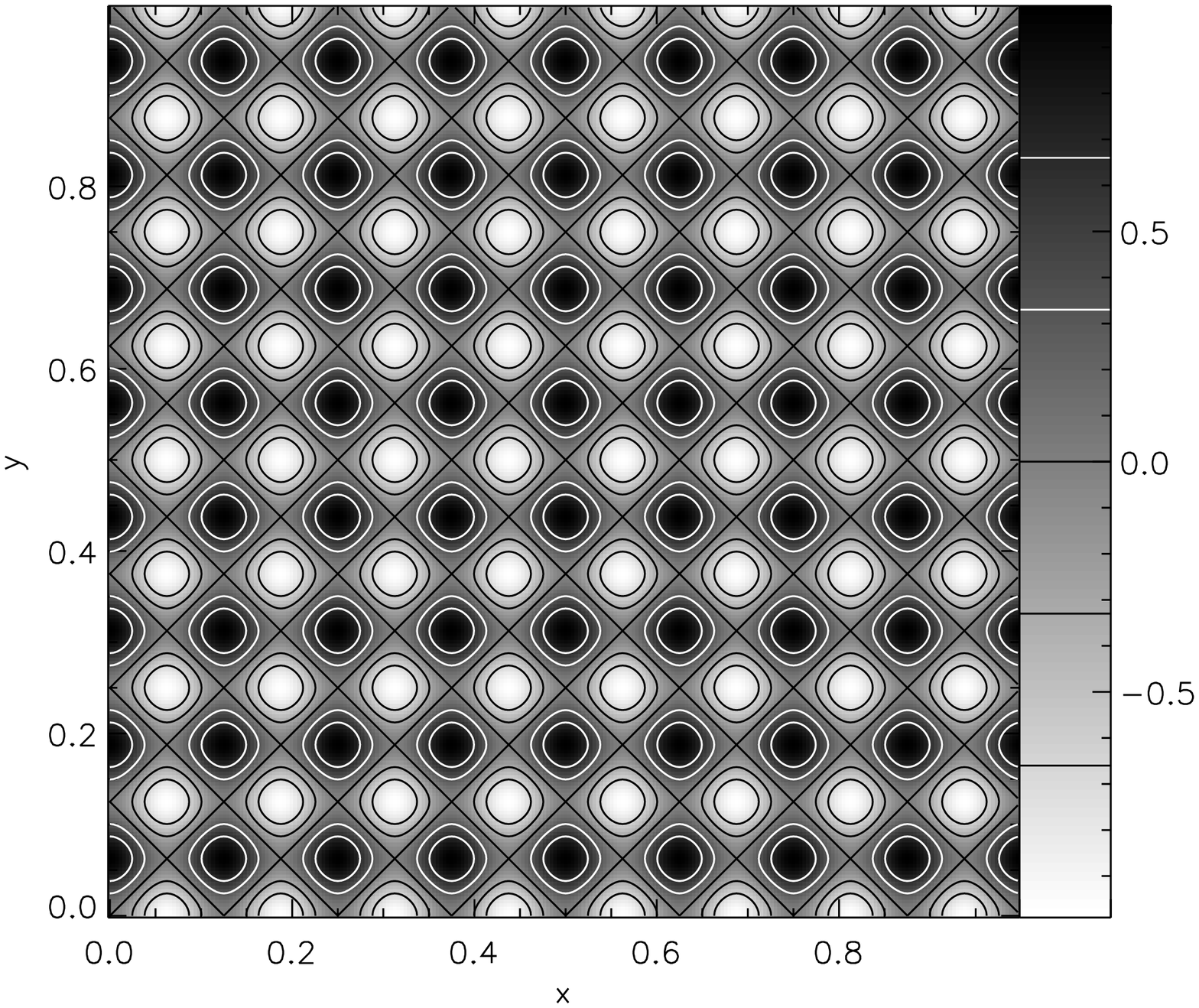, angle=90, width=6cm}
\caption{Examples of periodic data sets used to test the $\Delta$-variance
algorithm. The fBm structure on the left is characterised
by a spectral index $\zeta=3.0$. {\changed The chess board has a characteristic
structure size of about 0.13. The sine wave field on the right
consists of a single Fourier component at $k=8$, the corresponding
characteristic distance is $0.09$.}}
\label{fig_perdata}
\end{figure*}

\subsection{Test data sets}
\label{sect_testdata}

In order to test how the $\Delta$-variance reproduces specific structural
characteristics, we
have constructed a series of artificial data sets with known characteristics. 
They were either chosen to reproduce the typical self-similar scaling
behaviour measured in many astrophysical observations \citep{Combes}
or to contain pronounced artificial structures with a well known
size scale which should be clearly detected in the $\Delta$-variance
spectrum. {\changed
All test data sets were generated with an intrinsic resolution of
$128 \times 128$ pixels.}

\subsubsection{Periodic data}
\label{sect_pertestdata}

As a simple test bed where boundary effects play no role we started
with three types of periodic maps. The first type is provided
by the fractal structures of {\it fractional Brownian motion} (fBm)
as used by \citet{Bensch}. fBm structures are defined
by a power-law power spectrum $P(|\vec{k}|)\propto |\vec{k}|^{-\zeta}$
and random phases in Fourier space. We created fBms by the
Fourier transform of a Hermitian field with amplitudes
following the power spectral index $\zeta$ and random
phases. This procedure guarantees real values
and periodic maps. The structure analysis in terms of the
$\Delta$-variance should recover the power spectral
index of the fBm structure measuring a slope $\alpha=\zeta-2$
for this data set.

As periodic structure with a pronounced size scale
we use chess board like patterns where the number of fields on the
board is varied to change the size of the 
structures in the map relative to the map size.
Because the single chess fields are the
only structure in this data set, their size should appear
as prominent peak in the $\Delta$-variance spectrum.
This data set
gives a sharp definition of the characteristic length scale
in the spatial domain but contains a high contribution of
high frequency modes in the spatial frequency domain due to
the sharp edges of each field. Thus we have added as a third type of
test maps data fields provided by a single Fourier component
in both directions, i.e. the superposition of two 
orthogonal sine waves with the same period\footnote{All
maps are normalised to have a side length of 1 here.}, {\changed 
$f(\vec{r})=\cos(2\pi kx) +\sin(2\pi ky)$. In the case of }
a wavenumber
$k=1$ it can be regarded as an fBm with $\zeta \rightarrow \infty$.
{\changed The scale of the characteristic variation should be
clearly detected but} in contrast to the power spectrum we do not
expect a single sharp
maximum in the $\Delta$-variance spectrum, because the Fourier
transform of the $\Delta$-variance filter is a Bessel function with
pronounced side lobes.

{\changed The test data sets are all characterised by one free
parameter. For the fBms this is the spectral index $\beta$. 
For the chess board pattern and the sine wave field it is the
size of the characteristic structure or the dominant wavenumber,
respectively.}
Examples of the test data sets are shown in Fig. \ref{fig_perdata}.
The fBm structure used here
is characterised by a spectral index
$\zeta=3.0$ and a total variance of 1. The chess board 
example shows characteristic structure lengths between 0 and 0.24.
{\changed Its average size, integrated over
all possible angles is 0.13. The sine wave field shown here is
characterised by a wavenumber $k=8$ leading to a length of 0.09 for the
maximum variation.}

\subsubsection{Non-periodic data}
\label{sect_nonperdata}

As real astrophysical data are hardly periodic, 
tests for the treatment of boundary effects have to be performed
on non-periodic data. We use two types of data sets
to study these effects.

First we select subsets of larger (periodic) fBm structures
in the same way as introduced by \citet{Bensch}. The subsets cover
one quarter in length, i.e. $1/16$ in area, of the 
periodic fBm field so that they should hardly
retain any information about the large-scale periodicity. 
{\changed As the subsets are randomly chosen they typically
show sharp discontinuities at the edges. This seems to exclude
Fourier based methods for the analysis. In Sect.
\ref{sect_edgetreatment} we show, however, that the $\Delta$-variance
analysis can be extended to account for the discontinuities.}
Although it is not guaranteed that a subset has the same spectral index
as the whole fBm structure we will judge the value of the
$\Delta$-variance analysis based on the agreement of the determined
spectral index with the original fBm index because this approach
reflects the typical observational strategy that high
resolution observations are restricted to small
parts of a molecular cloud but they are used to derive
the general scaling behaviour of the cloud \citep[see e.g.
the IRAM keyproject ``Small-scale structure of pre-star-forming
regions'',][]{Falgarone}.

\begin{figure}
\centering
\epsfig{file=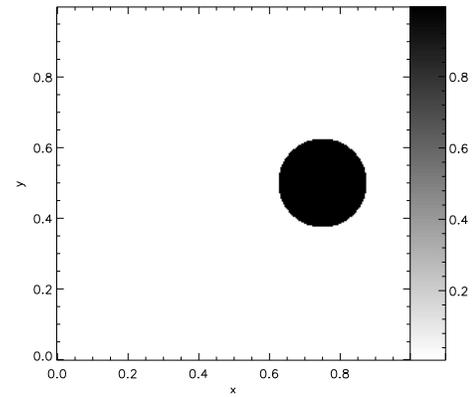, angle=90, width=6.0cm}
\caption{Example of a filled circle structure used to test
the $\Delta$-variance algorithm. The diameter of the circle 
is 0.25.}
\label{fig_nomcircleexample}
\end{figure}

As second non-periodic structure we use a filled circle on an
otherwise empty map. The dominant scale is the size of the
circle but as the $\Delta$-variance also measures the size
of the ``empty'' surroundings of the circle we expect 
considerable contributions from this area to the spectrum at
large lags. 
By adjusting the position of the circle with respect to the
map boundaries we can test the robustness of
the edge treatment in the non-periodic algorithm (in a periodic
treatment the $\Delta$-variance is independent of 
the position of the circle). {\changed The main free parameter of this
structure is the diameter $d$ of the circle. The average 
circle scale is given by $\pi/4 d$. We have not scanned
the distance to the map boundary as an additional free parameter,
but only performed a few tests for circles shifted by a 
integer multiples of the full diameter relative to the centre.}
An example with
a circle diameter of 0.25 is shown in Fig. \ref{fig_nomcircleexample}.

\section{Weighting in Fourier space}
\label{sect_edge}

\subsection{Edge treatment}
\label{sect_edgetreatment}

In the introduction of the $\Delta$-variance by \citet{Stutzki} 
we made extensive use of the Fourier transform, 
applying Eq. (\ref{eq_deltafourier}) to convolve the astrophysical
image with the $\Delta$-variance filter.
With the Fourier transform implicitly assuming data periodicity,
we introduce, however, steps at the edges of a map if the intensity
does not show the same value on both sides. 
Because step functions contain contributions at all spatial frequencies
they distort the $\Delta$-variance spectrum of the original structure.
\citet{Bensch} have shown that this effect
can lead to considerable errors for most astrophysical structures, 
where a periodic continuation is not possible.

The solution proposed there with the POINT and PIX algorithms was a
modification of the filter function when applied 
close to the boundaries of the maps. The filter is 
truncated so that it never stretches beyond the map edges.
To guarantee that the core and the annulus are still normalised to unity
for the truncated filter, both parts are multiplied with correction
factors depending on the remaining filter size. This edge treatment
has, however, the disadvantage that different parts of a map
are convolved with a different filter function so that the
computation of the $\Delta$-variance by Fourier transform
via Eq. (\ref{eq_deltafourier}) is no longer possible. 
In \citet{Bensch} we thus concluded that the $\Delta$-variance
should be rather computed in ordinary space.  This
approach is, however, slow compared to the computation in Fourier
space (several hours instead of a few seconds for maps containing
more than about 100$\times$100 pixels).

Here, we introduce a method that combines the improved edge
treatment with a computation in Fourier space. It is fast
and does not introduce artificial high-frequency contributions
{\changed by periodically wrapping discontinuities at the map
edges}. To obtain this behaviour the
method is set up to fulfil three conditions: {\bf i)} In the
convolution a fixed filter function is to be used so
that the convolution can be computed by multiplication in
Fourier space. {\bf ii)} To avoid effects of periodic
edge-wrapping, the filter contributions beyond the edges of
the map have to be truncated. {\bf iii)} The
{\changed normalisation of the filter discussed in Sect. 
\ref{sect_mathintro} has} to be fulfilled at each point.

These apparently conflicting requirements can be met using two
simple ideas. The truncation of the filter is substituted by a 
zero-padding of data and the corresponding error
in the filter normalisation is corrected by weighting
factors for the two filter contributions computed as a function
of the coordinates in the map.

Instead of truncating the filter when it extends beyond the
map edges we increase the map by zero-padding beyond the
edges up to the maximum filter size used. {\changed Then 
the extended map can be convolved with a fixed filter function, 
only providing zero contributions from outside
the original map. In this way no points from other periodically
wrapped parts of the image may then fall into the filter 
centred at any map position.} 
The error in the normalisation of the filter introduced
by this substitution can be computed from the convolution 
of the filter with 
an auxiliary map $w(\vec{r})$, which has a value 1 inside
the range of valid data and 0 in the zero-padded
region.
Because the $\Delta$-variance 
filter $\bigodot_l$ has to fulfil the two {\changed 
normalisation conditions for the core and the annulus 
(see Eq. \ref{fhat})}
\begin{equation}
\sum\sub{map} \bigodot\sub{{\it l}, core}(\vec{r}) =
\sum\sub{map} 
\bigodot\sub{{\it l}, ann}(\vec{r}) = 1 
\end{equation}
it has to be split into the positive and negative filter parts
for the computation of the normalisation
errors. The sums extend over the map of valid data.
In total four convolution integrals need to be  
evaluated to compute the $\Delta$-variance
\begin{eqnarray}
G_{l, {\rm core}}(\vec{r}) &=& f\sub{padded}(\vec{r}) * 
\bigodot_{l, {\rm core}}(\vec{r'}) \nonumber \\
G_{l, {\rm ann}}(\vec{r}) &=& f\sub{padded}(\vec{r}) * 
\bigodot_{l, {\rm ann}}(\vec{r'}) \nonumber \\
W_{l, {\rm core}}(\vec{r}) &=& w(\vec{r}) * 
\bigodot_{l, {\rm core}}(\vec{r'}) \nonumber \\
W_{l, {\rm ann}}(\vec{r}) &=& w(\vec{r}) * 
\bigodot_{l, {\rm ann}}(\vec{r'}) 
\end{eqnarray}
where $f\sub{padded}$ stands for the map with the additional
zero-padded boundary region. {\changed As we use fixed filter functions
and all involved functions vanish at the map edges,} the convolution
integrals can be easily computed in Fourier space involving
a fast Fourier transform and a map multiplication. {\changed Although
the map treated in this way is larger than the original map by up to
a factor three in each direction, the Fourier transform using an
algorithm like FFTW, which can work on arbitrary map sizes, is still
considerably faster than any convolution in ordinary space.}

The full map convolved with the $\Delta$-variance
filter truncated at the map edges is then
\begin{equation}
F\sub{l}(\vec{r}) = {G\sub{{\it l}, core}(\vec{r}) \over
W\sub{{\it l}, core}(\vec{r})}
 - {G\sub{{\it l}, ann}(\vec{r}) \over
W\sub{{\it l}, ann}(\vec{r})}
\end{equation}
It is only defined where the normalisation parameters $W\sub{{\it l}, core}$
and $W\sub{{\it l}, ann}$ are both different from zero. 

In the computation of the $\Delta$-variance spectrum
one has to take into account the reduced significance of 
the data values in the convolved maps produced by the fact that
the applied filter becomes more and more distorted relative to
the optimum filter when it is truncated. Using the normalisation
parameters of the truncated filters as a measure for their significance
one can add a significance weighting to the $\Delta$-variance
analysis. We define the $\Delta$-variance 
no longer as the variance of the convolved map but weight {\changed the
map points by the significance of the filter applied to
compute the value at each point when computing the variance}
\begin{equation}
\sigma_\Delta^2(l) = { \sum\sub{map} (F\sub{l}(\vec{r}) - \langle F\sub{l}
\rangle )^2 W\sub{{\it l}, tot}(\vec{r}) \over \sum\sub{map}
W\sub{{\it l}, tot}(\vec{r}) }
\end{equation}
with
\begin{equation}
W\sub{{\it l}, tot}(\vec{r})=W\sub{{\it l}, core}(\vec{r})
W\sub{{\it l}, ann}(\vec{r})
\end{equation}
The sum covers the whole (extended) convolved data field.
The definition of the significance function as the product of
both normalisation factors is somewhat arbitrary but reproduces
the desired behaviour that changes in the positive and negative
part of the filter contribute equally.
We have tested different powers of the product but found the best
agreement with the theoretical behaviour in fBm data sets
for an exponent of just unity.

\begin{figure}
\epsfig{file=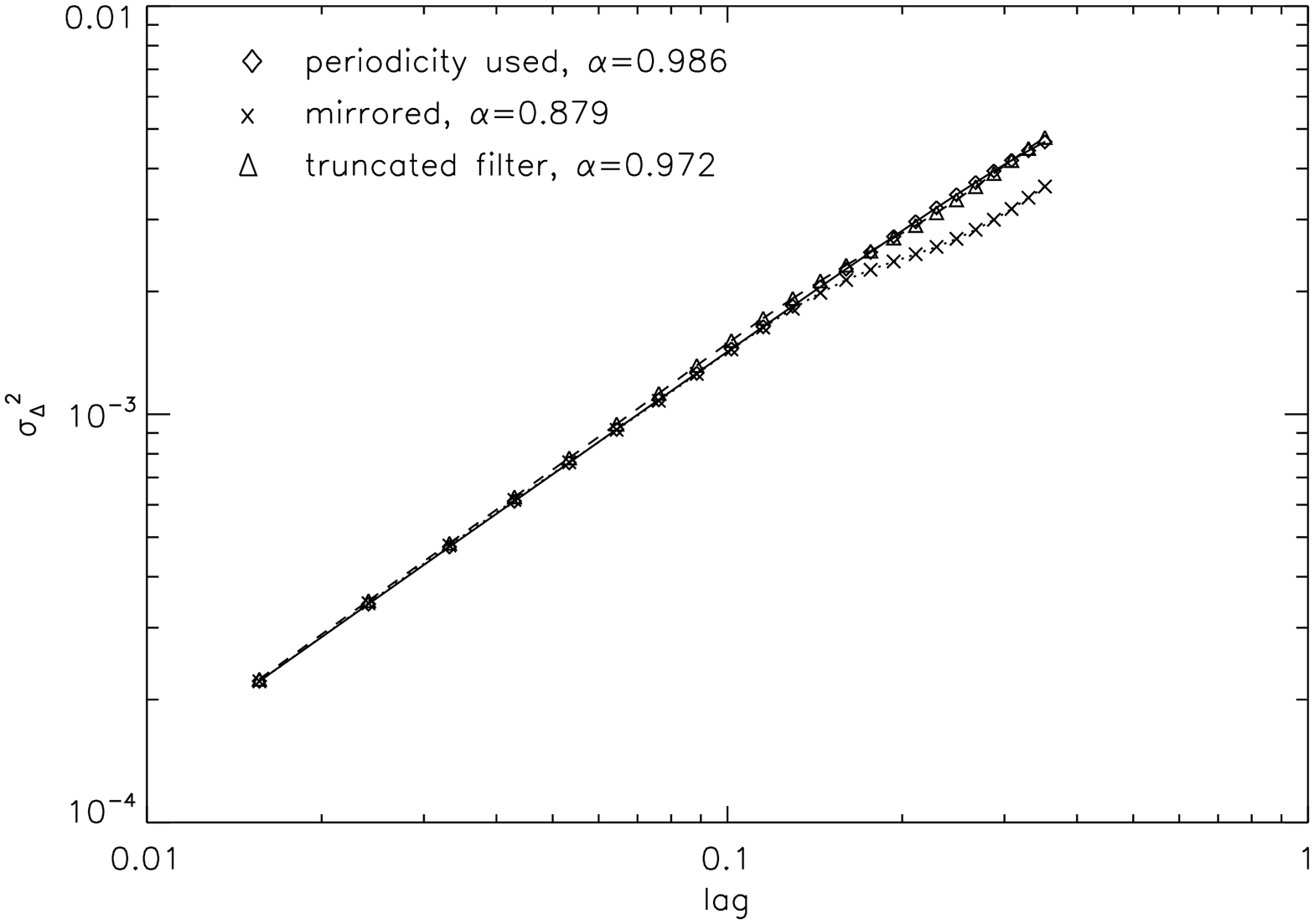, angle=90, width=\columnwidth}\vspace{0.3cm}
\epsfig{file=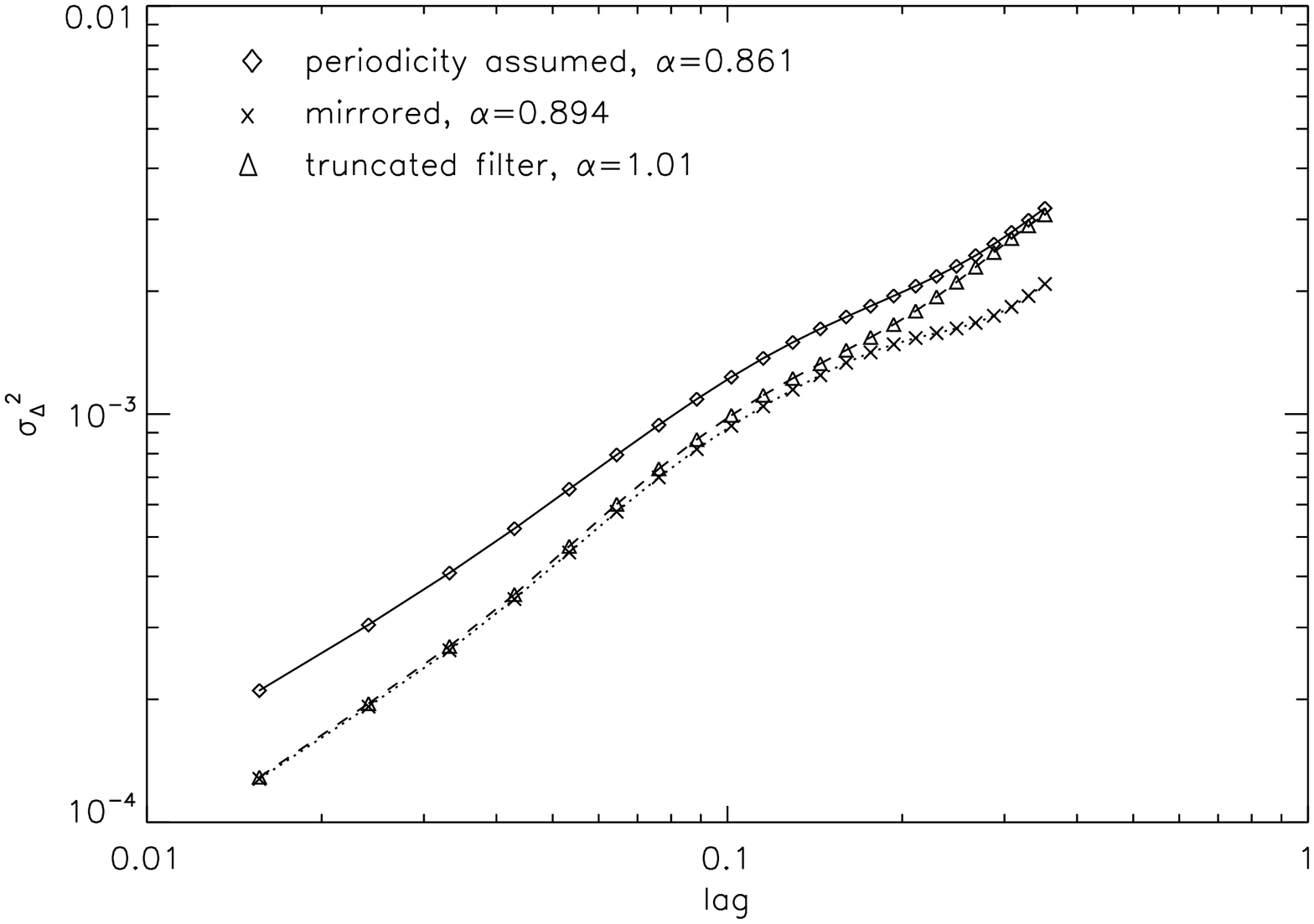, angle=90, width=\columnwidth}
\caption{$\Delta$-variance spectra computed with three different
types of the edge treatment for {\changed one }periodic fBm structure (upper plot)
and {\changed one }non-periodic subset from a fBm structure (lower plot). The fBms are
characterised by a spectral index $\zeta=3.0$ so that the
$\Delta$-variance should show a slope $\alpha=1.0$. The measured
slopes are indicated for each graph in the plot.}
\label{fig_edgeexample}
\end{figure}

Figure \ref{fig_edgeexample} demonstrates the effect of the edge
treatment in the example of an fBm structure which is periodic
and for a non-periodic sub-map from a larger fBm structure.
We have used three different ways to compute the $\Delta$-variance.
First we assume that the maps are periodic, neglecting
all wrap-around effects. For the periodic fBm this is, of course,
the best assumption which should reproduce exactly the
properties of the power spectrum used to generate the fBm. 
It is, however, rarely useful when dealing with observed data
as they are in general not periodic. The second approach creates
periodicity by mirroring the map along both axes as discussed by
\citet{Stutzki} so that a larger
periodic map is produced and wrap-around effects can be
neglected. This approach, however, still results in discontinuities
in the first derivatives at the mirror axes. The third approach 
uses the truncated filter as described above.

For the periodic fBm structure we find a very close agreement of the
$\Delta$-variance using the truncated filter with the theoretical
value given by a slope $\alpha=1$. The mirror continuation results in 
an apparent
reduction of the amount of large-scale structures in the map as
they are partially assigned to larger modes only present in the
map extended by mirroring. Thus the $\Delta$-variance slope
is systematically underestimated. For the non-periodic structure,
the assumption of periodicity {\changed results in strong deviations
from the power-law behaviour on small scales due to artificial 
high-frequency contributions from the edges.
The mirroring shows again an underestimate of the power spectral
index on large scales,} and only the use of the truncated
filters results in a good reproduction of the expected value for the slope.

\begin{figure}
\epsfig{file=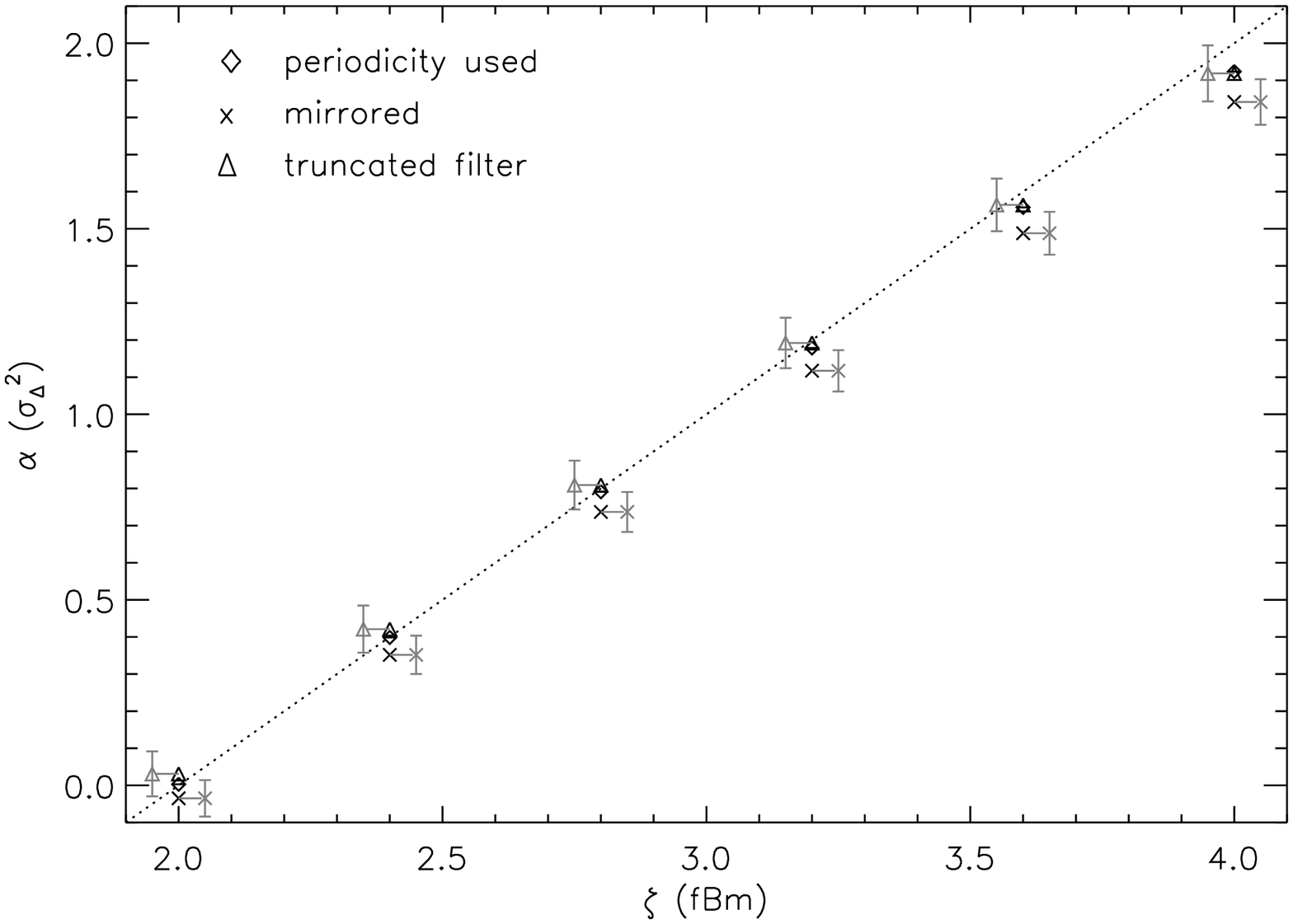, angle=90, width=\columnwidth}\vspace{0.3cm}
\epsfig{file=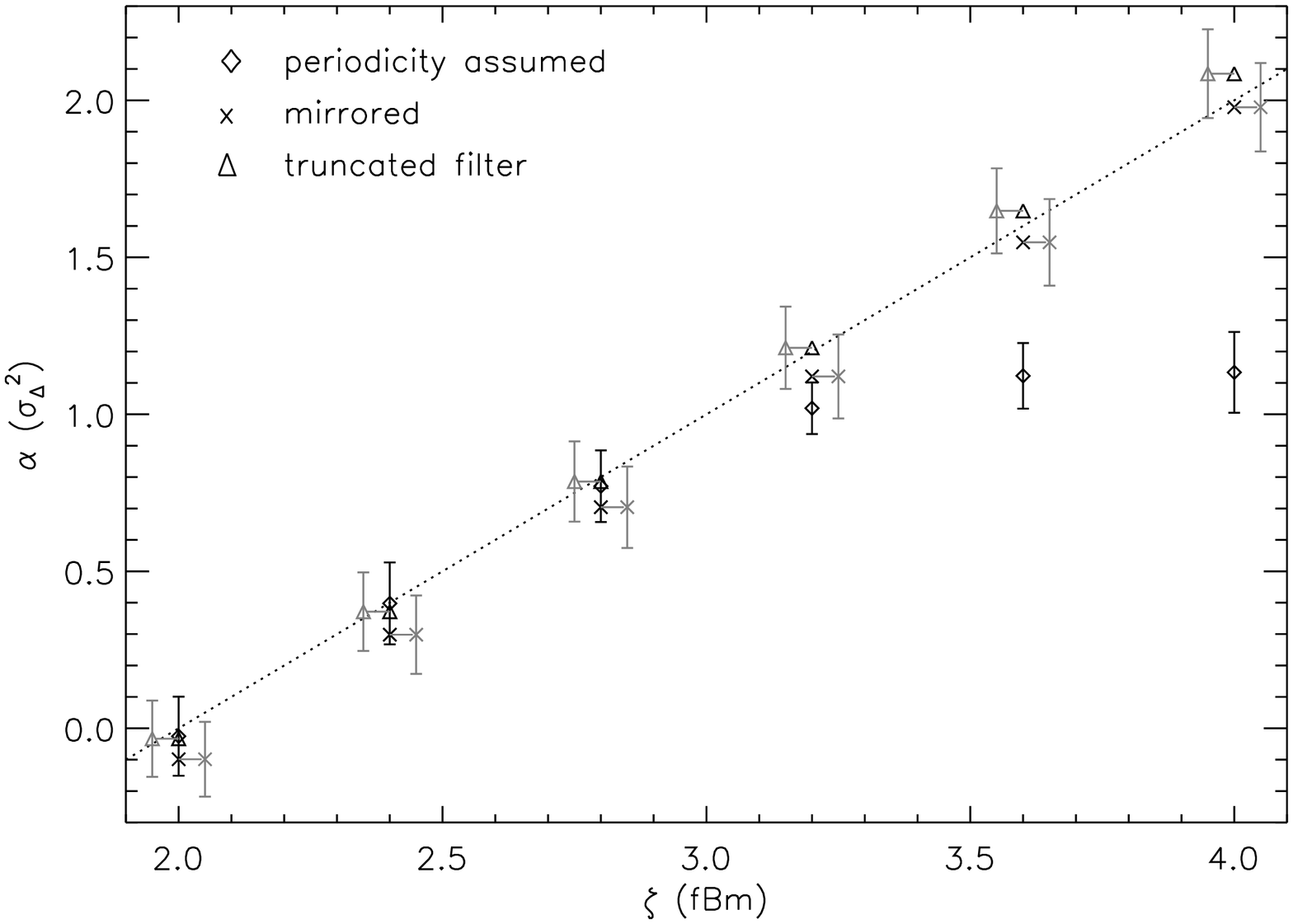, angle=90, width=\columnwidth}
\caption{Average and standard deviation of the spectral indices
of the $\Delta$-variance spectra determined for a set of fBm
structures (upper plot) and for a set of sub-maps from fBm structures
(lower plot) depending on the spectral index of the fBms.  All spectra
are computed at the same power spectral indices $\zeta=2.0, 2.4, 2.8,
3.2, 3.6,$ and 4.0, but for a better visibility, the standard
deviations for the truncated filter and for the mirror continuation are
displaced in the plot relative to these values. In the case of the
periodic analysis of the fBm maps the error bars giving the standard
deviation of the statistical samples are practically zero. The dotted line
indicates the theoretical index for infinitely large fBm structures
$\alpha=\zeta-2$. }
\label{fig_edgesystem}
\end{figure}

{\changed In Fig. \ref{fig_edgeexample} we plot the results obtained 
just for one realization
of an fBm and for one sub-map from a large fBm. But the exact shape
of the computed $\Delta$-variance spectra can depend on the distribution
of the random phases within the fBm and it will certainly depend on
the selection of the sub-map within an fBm. Thus
we have repeated the computation for a set of maps in Fig. \ref{fig_edgesystem}.}
Analogously to the statistical treatment by \citet{Bensch} we vary
the spectral index of the fBm structures and chose randomly 
30 different fBms or fBm sub-maps and determine their $\Delta$-variance
spectra. {\changed We do not display all spectra, but only the resulting distribution
of slopes $\alpha$.
The Fig. shows the average $\Delta$-variance slope and the spread of
measured slopes} as a function of the spectral index $\zeta$ using 
the three different edge treatments. {\changed All slopes are computed
from a power fit covering the full data range plotted in Fig. \ref{fig_edgeexample}.}
%The standard deviation in 
%each set is plotted as error bar. Because they overlap in all cases
%we plot the values for the filter truncation and for the mirroring at 
%shifted $\zeta$-coordinates, so that they can be easily distinguished by eye.
An optimum treatment
should reproduce the relation $\alpha=\zeta-2$ indicated by the
dotted line in Fig. \ref{fig_edgesystem}.

For the periodic fBms we find that the $\Delta$-variance spectra from
the truncated filter show about the same spectral indices as
the periodic treatment. 
The error bars in the periodic treatment are zero because the
$\Delta$-variance spectrum is independent of the exact phase
distribution and thus identical for each map of the sample.
The standard deviation of the slopes in the truncated-filter
treatment is always about 0.08 here. The $\Delta$-variance spectrum
underestimates the power spectral index by 0.03 at $\zeta=3.6$ and
by 0.08 at $\zeta=4$ because the theoretical value is only reached
for infinitely large maps and the deviations grow when approaching
the asymptotic limit of $\alpha=4$ (see Stutzki et al. 1998). The
$\Delta$-variance spectrum computed for the mirror-continuation
of the map always underestimates the spectral index by 
about 0.1.

In the case of non-periodic fBm sub-structures the periodicity assumption
clearly fails. {\changed With this approach the measured slope saturates
at about $\alpha=1.1$.} The use of the simple $\Delta$-variance without filter
truncation or mirror continuation provides wrong results
at power spectral indices above about $\zeta=2.9$. {\changed
This is due to unavoidable discontinuities
at the submap edges which result in high-frequency contributions
in the periodic treatment resulting in too shallow $\Delta$-variance spectra.}
In contrast,
both the mirror-continuation and the filter truncation provide
a reasonable measure for the actual map structure for all spectral indices.
{\changed Both methods are hardly affected by the discontinuities
at the sub-map edges.}
The mirror-continuation always underestimates the spectral index by 
about 0.1 (except at $\zeta=4.0$). The filter truncation method
reveals the correct spectral index
for $\zeta \le 3.0$ and overestimates it by 0.05 at $\zeta=4.0$.
Regarding the typical error bars of 0.15, the systematic errors are,
however, lower than the scatter between different
fBm realizations with the same spectral index.

The same tests were repeated for map sizes ranging
from $32^2$ to $256^2$ pixels. In agreement with the 
studies by \citet{Bensch} we found no systematic changes
in the $\Delta$-variance slopes exceeding 0.03 when changing
the map size but an increase of the error bars from 0.15
at map sizes of $128^2$ to 0.25 at a size of $32^2$. 
This higher uncertainty for smaller maps prevents any
significant conclusion from maps spanning less than $\approx 30$
pixels.

\begin{figure}
\epsfig{file=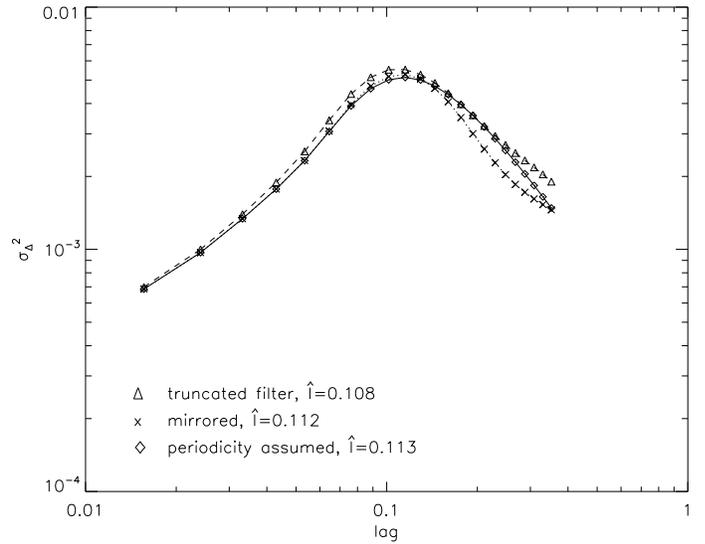, angle=90, width=\columnwidth}
\caption{$\Delta$-variance spectra for the filled-circle map
where the circle has a diameter of 1/8 of the map size and is
shifted by three diameters from the centre of the map. The
values $\hat{l}$ indicate the lag of the measured peak in the
spectrum.}
\label{fig_circleexample}
\end{figure}

To demonstrate the influence of the edge treatment on the $\Delta$-variance
spectrum of an object with a pronounced size scale
we show in Fig. \ref{fig_circleexample} the spectra computed
for the map containing the filled circle with a diameter of
1/8 of the map size. The peak of the $\Delta$-variance spectra at 0.11
falls below the circle diameter of 0.125 but slightly above
the average distance between two points on the rim of the circle {\changed
which is about 0.10. The somewhat higher value} is probably due to the
contributions from the ``empty'' environment of the circle at 
large lags, also resulting {\changed in a decay of the
$\Delta$-variance spectra at large lags which is shallower than the
$l^{-2}$-characteristics of uncorrelated structures.}. These contributions
are dispersed over a relatively wide range
of scales corresponding to the different distances to the
map boundary in a non-periodic treatment and to the distances
to the next circle in a periodic interpretation of this structure.
As these variations are mainly assigned to lags exceeding the
map size in the periodic treatment, the two curves for the
periodic treatment show somewhat lower $\Delta$-variance values
within the map than the filter truncation method where the
``empty'' region is constrained by the map size.
Nevertheless, the total differences between $\Delta$-variance
spectra using the different edge treatment methods are relatively
small, so that either method seems to be justified for this case.

% No important conclusion - thus removed: ossk
% When the circle is shifted to the centre of the map, the
% periodic treatment and the mirror-continuation obviously agree
% among each other and with the spectrum shown here for the
% periodicity assumption. The spectrum computed with the
% truncated filters is somewhat more  increased at large lags
% being a factor 1.5 higher than the other two spectra  at
% the largest lag shown here.

\subsection{Generalisation for data with varying reliability}

The concept of weighting the $\Delta$-variance computation by
different filter significance values can be generalised to
deal with data, where the data points in a map are as well
characterised by a variable data reliability. This applies e.g. to maps
where not all points are observed with the same integration time
so that they show a different noise level. The inverse noise
RMS is an indicator for the significance of the data at
different points. Many other observational effects may lead to a
similar variation in the data reliability across the map.
As long as the reliability can be expressed as a significance number 
$w\sub{data}(\vec{r})$ between 0 and 1 all such maps may be analysed
within the concept outlined here. 
The same equations as discussed in the filter truncation are to be 
applied, but the auxiliary weight map $w(\vec{\vec{r}})$ does
no longer consist of the values 1 inside and 0 outside of the
original map. It rather contains
the significance values 
$w\sub{data}(\vec{r})$ ranging continuously from 0 to 1.  The
weighting factors in the $\Delta$-variance computation
$W\sub{{\it l}, tot}(\vec{r})$ then contain the integrated significance
of the filter-convolved data at each point.
 
With this generalised concept, the $\Delta$-variance analysis can be applied
to arbitrary two-dimensional data sets. They must be 
projected onto some regular grid but they do not need to contain regular
boundaries as the corresponding ``empty'' grid points only have to
be marked with a zero significance. Varying noise or other changes
in the data reliability can be expressed in the significance
function $w\sub{data}(\vec{r})$ which has to be constructed for
each data set. The only remaining requirement for the applicability of the
$\Delta$-variance is the sufficiently large spatial dynamic range
in the data. The criterion of at least 30 pixels in each direction
for reasonable error bars of the $\Delta$-variance spectrum 
discussed above has to be extended in the case of a low data significance.
{\newchanged In the appendix we present measurements of the dynamic
range over which the slope of fBms can be reliably determined
in the case of noisy data. We find as a rule of thumb that the minimum 
map size has to be increased by one over the average data significance.}
In paper II we will apply the $\Delta$-variance analysis to observed data
with irregular boundaries and a spatially varying significance.

\section{Filter optimisation}
\label{sect_filter}

\subsection{The $\Delta$-variance filter function}

All examples given above were computed with the fixed filter function 
of a French hat with a diameter ratio between the annulus and the core
$v=3$. An obvious question is whether we can
improve on the $\Delta$-variance by using a different diameter ratio
and/or a different filter function. Due to its
discontinuity in the normal space the French hat has high
frequency lobes in Fourier space. Alternative approaches should
use smoother functions in ordinary space to obtain a better
confinement in Fourier space. As a smooth example we implemented
a ``Mexican hat'' consisting of two Gaussian functions:
\begin{eqnarray}
\bigodot_{l, {\rm core}}(\vec{r}) &=& {4 \over \pi l^2}
\exp\left(\vec{r}^2 \over (l/2)^2 \right) \\
\bigodot_{l, {\rm ann}}(\vec{r}) &=& {4 \over \pi l^2 (v^2-1)}
\left[ \exp\left(\vec{r}^2 \over (vl/2)^2\right)
- \exp\left(\vec{r}^2 \over (l/2)^2\right)\right]\nonumber
\end{eqnarray}
where $l$ is the size of the filter and $v$ is the diameter ratio
between the annulus and the core of the filter as defined in
Sect. \ref{sect_mathintro}. The choice of the Gaussians guarantees
that the filter gives the best simultaneous confinement
in ordinary space and in Fourier space not showing any 
side lobes in either of them.

As the opposite extreme we have also tested a filter measuring
the difference between one point in the map and all points displaced
by the sharp distance $l$ relative to this pixel. The corresponding
$\Delta$-variance then measures basically the structure function of the
map \citep[see e.g.][]{Miesch94, Frick}. However,
because those tests only confirmed the results from \citet{OML}
and Frick et al. (2001) that the structure function is relatively
insensitive to distortions of the power spectrum on
particular scales we excluded this filter from the following
studies.

Thus, we restrict ourselves here to the Mexican- and the
French-hat filter, where we vary in both cases the diameter
ratio $v$. In this way we test both the influence of the
general filter shape and of the ratio between core and annulus
for each filter on the computed $\Delta$-variance spectra.

\subsection{The effective filter length}
\label{sect_effectivelength}

Taking into account 
the finite width of the core and the annulus of both filter
functions it is not obvious on which scale variations are
actually measured when using a filter with size $l$. Here,
we compute this scale based on the geometrical properties of the filter.

In the French-hat filter we measure the average distance
between a point in the core and a point in the annulus, providing the
average scale on which a structure variation in the map should be
measured. For the Mexican-hat filter, the computation of the average
distance includes the additional weighting of each distance by the product of
the positive and negative filter values. This reflects the effect
of the convolution of a map with this filter.

\begin{figure}
\epsfig{file=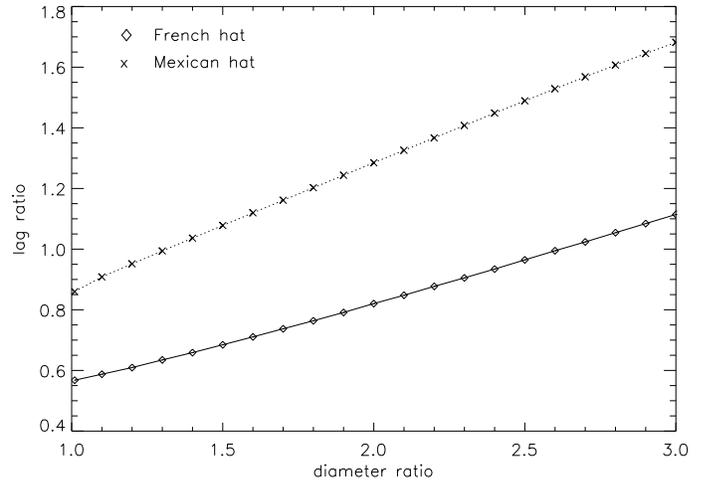, angle=90, width=\columnwidth}
\caption{Factor translating the filter core diameter $l$ into the
average distance measured by the filter as a function of the diameter
ratio between annulus and core of the filter. The average distance is
computed by a double integral over the core and the annulus of the
filter.}
\label{v-abstand}
\end{figure}

Figure \ref{v-abstand} shows the resulting effective filter length relative
to the filter size $l$ as a function of the diameter ratio $v$ for
the French and the Mexican hat.
The length scale traced by the filters is approximately a
linear function of the diameter ratio between the core and the annulus
of the filter. {\changed Using a least-square fit we obtain the coefficients}
\begin{equation} 
{l\sub{eff} \over l} =\left\{\displaystyle
0.29 v+0.26 \quad  {\rm for\; the\; French\; hat} \hfill
\atop \displaystyle
0.41 v+0.46 \quad {\rm for\; the\; Mexican\; hat.} \hfill \right.
\end{equation}

The original $\Delta$-variance definition using a French hat with
$v=3$ gives an effective length of 1.12 times the core diameter
$l$. Thus all scales computed previously with that filter 
should be shifted by the factor 1.12. This is only a small correction, not
changing the conclusions in any of the papers that have used the
$\Delta$-variance so far. In all plots shown in this paper, we use the
effective length as the lag of the $\Delta$-variance 
to allow a direct comparison of the spectra independent of
the filter used. 

\subsection{Filter evaluation}

The optimum filter to be used in the $\Delta$-variance analysis has to
fulfil two criteria: the correct detection of pronounced size scales in
the maps and the exact determination of the scaling exponents of the
contained structures.

\subsubsection{Scale detection}
\label{sect_scaledetect}

In the detection of pronounced scales the
maximum of the $\Delta$-variance spectrum should fall onto the correct
lag corresponding to the structure size.  Moreover, the signature of
the pronounced scale in the $\Delta$-variance spectrum should be as
sharp as possible with a high contrast relative to other scales. 
As test images we used the chess board field, the sine wave field, and
the filled circle field.

\begin{figure}
\epsfig{file=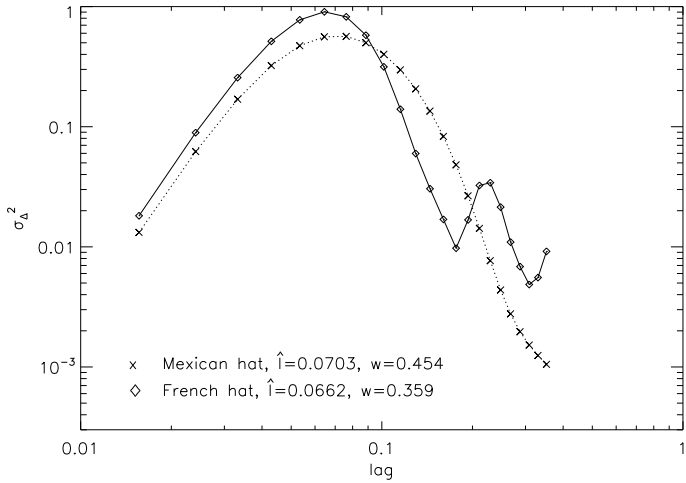, angle=90, width=\columnwidth}
\caption{$\Delta$-variance spectra computed for the sine wave
field with a wavenumber $k=8$. The filter truncation is
applied in both cases for the edge treatment. {\changed The peak scale
is given as $\hat{l}$, 
the logarithm of the ratio between
upper and lower lag where the $\Delta$-variance has dropped to
$1/2$ of the peak value is given as $w$.}}
\label{fig_sinedeltas}
\end{figure}

{\changed To illustrate the behaviour we plot the results for the sine wave field.
Fig. \ref{fig_sinedeltas} shows the $\Delta$-variance spectra measured
for a field with $k=8$ by the French and the Mexican-hat filters.}
In both cases the
filter truncation method is used. The dominant scale is 
detected as a peak in the $\Delta$-variance spectra, the peak position
at about 0.07 is somewhat lower than the scale
of the maximum variation $1/(k \sqrt{2})=0.088$. We found this small shift by about 20\,\%
in all spectra. When using the effective filter length, the peak position
is very constant independent of the filter type and its diameter
ratio.

{\changed On scales above the peak, the spectra obtained for the two filters
deviate considerably. The French-hat filter produces ripples at large
lags. From the positions of these ripples we
find that they reflect the side lobes of the Bessel function,
representing the Fourier transform of the French-hat filter.
This must not be misinterpreted as a detection of large-scale
structure in the map. The side lobes detect
the single Fourier amplitude from $k=8$ also at other
effective filter sizes. This can be seen when changing the
diameter ratio $v$ in the filter.} The peak remains at the same
position but the ripples in the $\Delta$-variance spectrum move
corresponding to the changes in the side lobes of the Bessel
function. Decreasing $v$ relative to the value of 3.0 results in a
steeper decay above the peak and an increased number of ripples at
large lags. When reducing the diameter
ratio $v$ for the Mexican-hat filter the decay above the peak also
steepens and we obtain some flattening at the largest lags in the map.
In general we can either achieve the sharp peak and the artificial
structures at large lags by the French-hat filter or the
broader peak without ripples by means of the Mexican hat. 

{\changed Because the field is periodic we can also apply the periodic continuation
method here.} The equivalent figure shows the same shape of the peak but more
pronounced ripples with deeper minima on large scales for the French
hat filter and a somewhat steeper decay {\changed for the Mexican
hat. In this case the $\Delta$-variance spectrum simply represents 
the Fourier transform of the filter function because the structure
contains only a single Fourier component. For integer wavenumbers the periodic
continuation is identical to the mirror continuation so that we obtain
the same spectra} except for a
small discretisation error from a single pixel row which is treated
different when mirroring or continuing periodically. In general the
edge truncation always produces a strong and the mirror-continuation a
weak smearing of the French-hat $\Delta$-variance ripples at large lags relative
to the periodic continuation.

Corresponding results for the chess board map show almost the same
curves as the sine wave field 
at all lags above the peak in the $\Delta$-variance spectra.
But they show a flatter rising part at smaller lags.  The
additional high-frequency contributions from the edges of the chess
fields increase the $\Delta$-variance at small lags {\changed
leading to a slope shallower than 1 there}. 
% so that the small-scale slope is changed from a value close
% to 4 for the sine wave field to 0.8. 
The situation is somewhat different for the
circle map. The strong contributions from the ``empty'' area at
large lags visible in Fig. \ref{fig_circleexample} suppress 
differences from the different filter shapes
in the spectrum. The spectra obtained from both filter types
look very similar.

\begin{figure}
\epsfig{file=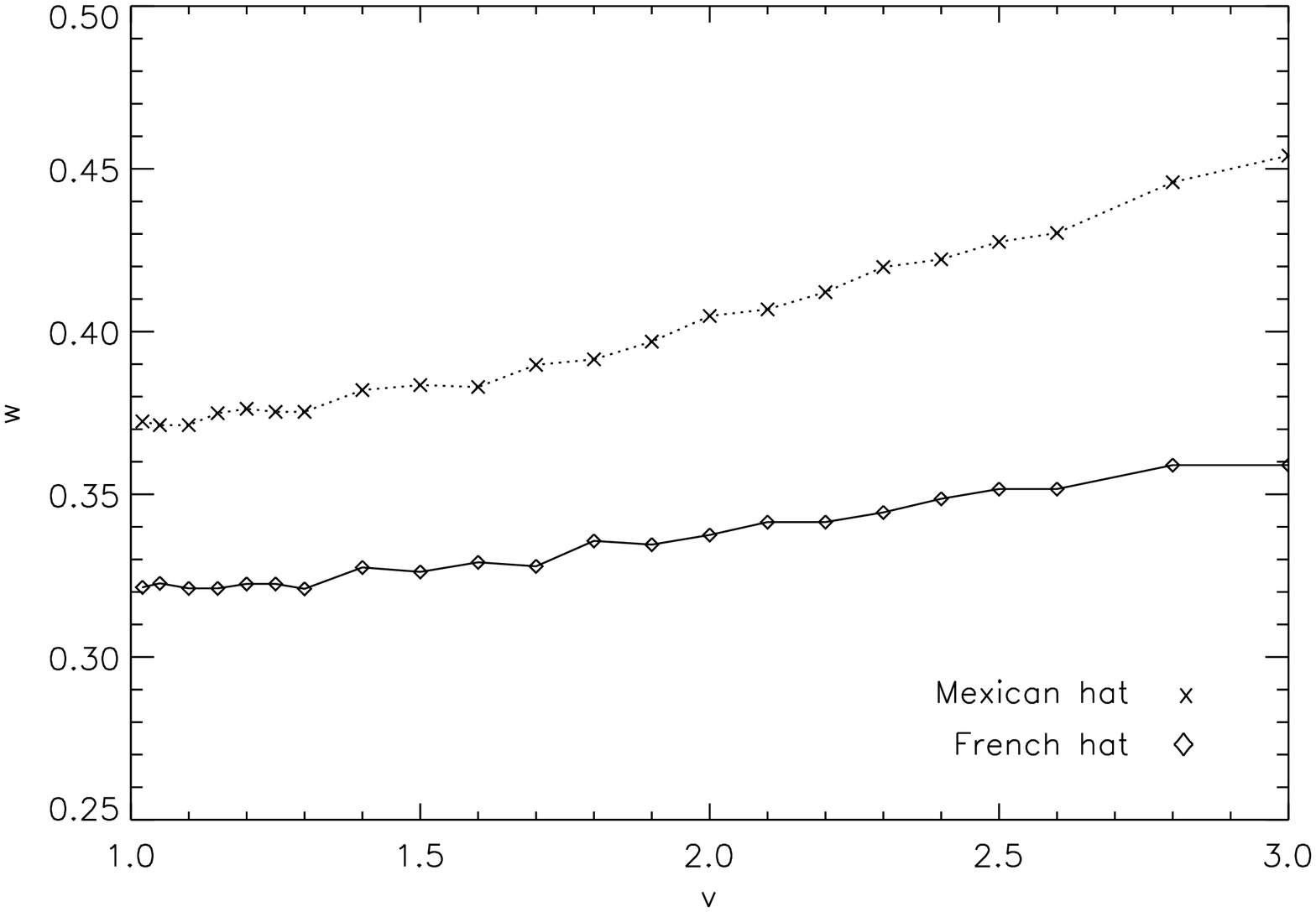, angle=90, width=\columnwidth}\vspace{0.3cm}
\epsfig{file=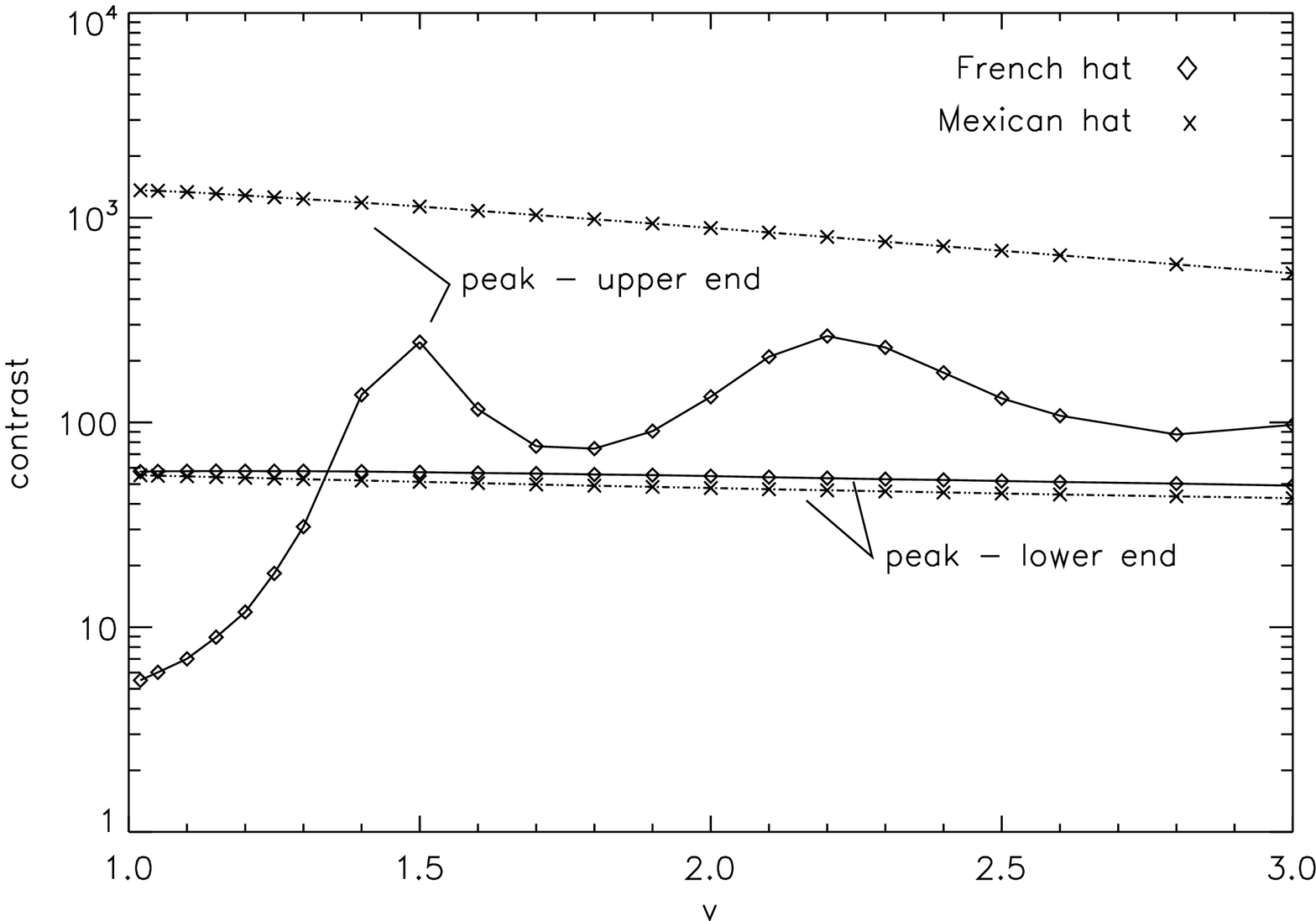, angle=90, width=\columnwidth}
\caption{Logarithmic FWHM of the $\Delta$-variance peak and
contrast between the peak value and the values at $l=0.014$
and $l=1/3$ for the sine wave map analysed in Fig.
\ref{fig_sinedeltas}.}
\label{fig_sinecontrast}
\end{figure}

To quantify the influence of the selection of the filter shape and
the diameter ratio on the detection of the dominant scale
length we measure the sharpness of the $\Delta$-variance
peaks shown in Fig. \ref{fig_sinedeltas}. 
We use the {\changed width of the peak $w$, given as the logarithm of the ratio
between upper and lower lag where the $\Delta$-variance drops to
$1/2$ of the peak value, and its contrast relative
to the values at lags which are either
small or large compared to the peak position, i.e. relative to
the first and the last point in the $\Delta$-variance spectrum.}
The behaviour of both parameters is shown in Fig. \ref{fig_sinecontrast}
as a function of the filter diameter ratio $v$. The upper
plot shows the logarithmic width of the peak; the lower plot the
contrast of the peak relative to values at much larger and much
smaller lags. From the figure it is obvious that we cannot achieve
a minimum width and maximum contrasts simultaneously with the same filter, so that
some balance has to be found.
The contrast with respect to smaller lags hardly varies
with filter type and diameter ratio but the contrast relative to large lags is
drastically changed. The French-hat filter always produces a narrower
peak but the decrease of the peak width towards lower
$v$ ratios is accompanied by a deterioration of the contrast with
respect to large lags. The Mexican-hat filter shows a continuous
improvement of both quantities towards lower ratios but gives a 
somewhat broader peak.
The slight increase of the contrast with respect to the lower end
of the spectrum towards lower $v$ values for both filter types
indicates that low diameter ratios result in a somewhat longer
dynamic range below the dominant peak where the $\Delta$-variance
spectrum follows a power law.

Comparing the different edge treatment approaches in 
corresponding plots shows that the mirror-continuation always
produces about the same peak width and a somewhat better contrast 
than the filter truncation method because the latter 
introduces some flattening in the $\Delta$-variance spectra at 
the largest lags.

\begin{figure}
\epsfig{file=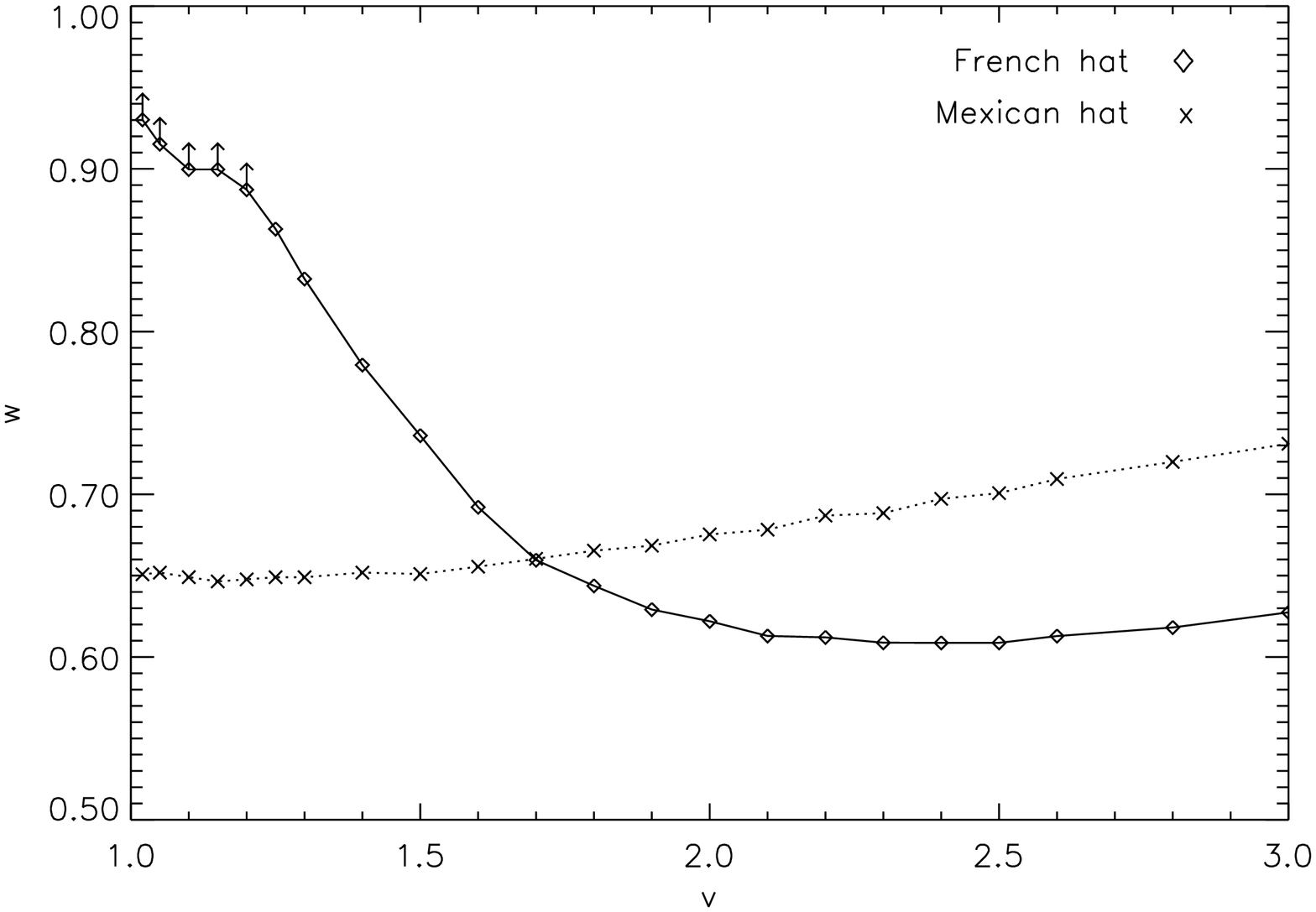, angle=90, width=\columnwidth}\vspace{0.3cm}
\epsfig{file=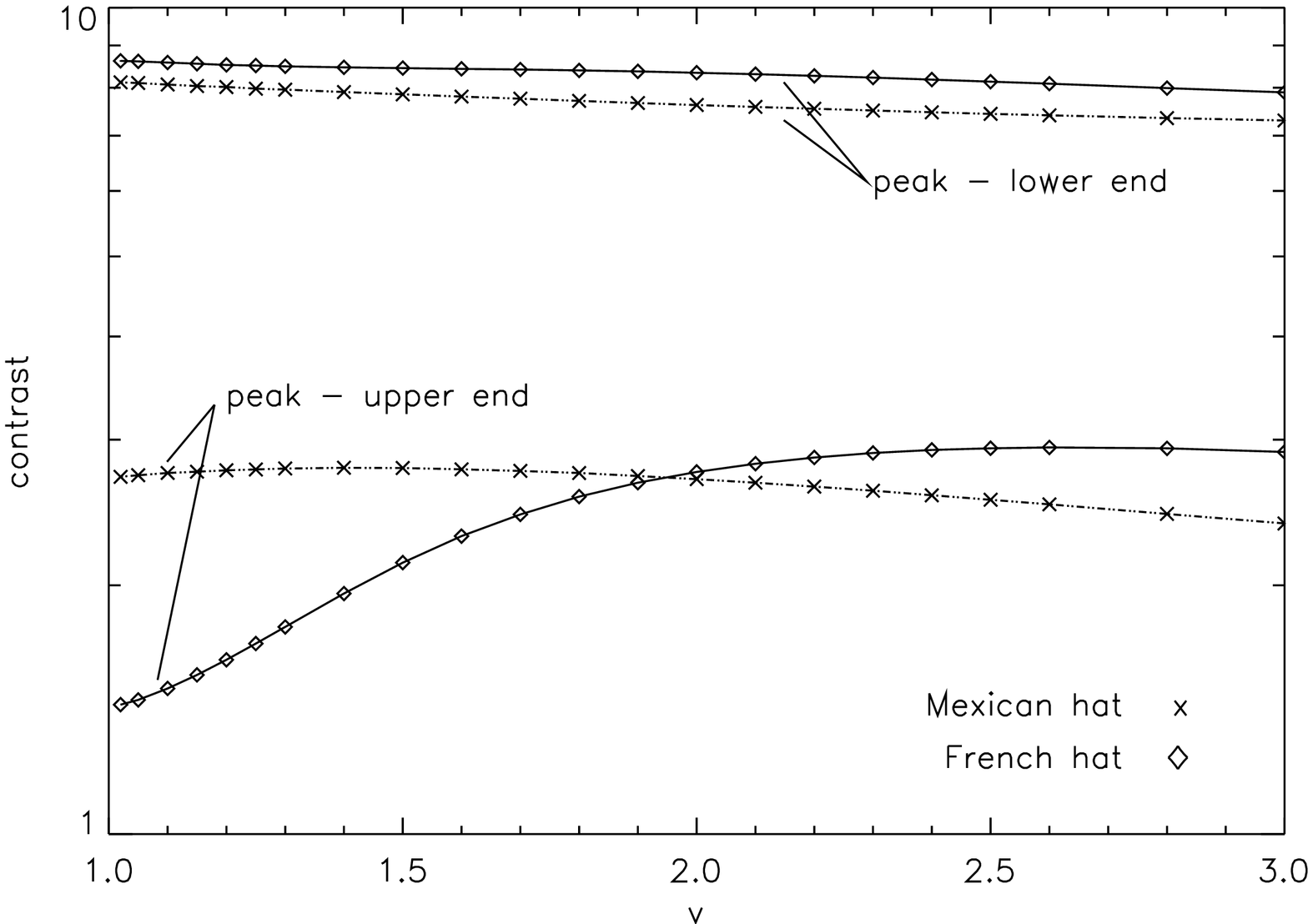, angle=90, width=\columnwidth}
\caption{Logarithmic FWHM of the $\Delta$-variance peak and
contrast between the peak value and the values at $l=0.014$
and $l=1/3$ for the filled circle map discussed in Fig.
\ref{fig_circleexample}. The widths measured for the French hat
at $v<1.25$ are lower limits because the spectrum
did not fall below half of the peak value at the
upper end of the spectrum.}
\label{fig_circlecontrast}
\end{figure}

A very similar behaviour is also observed for the chess board structure.
In contrast, the filled-circle map with its large-scale contributions
shows a different behaviour demonstrated in Fig. \ref{fig_circlecontrast}. 
Only the contrast relative to the $\Delta$-variance values at small lags
shows the same small improvement for both filters towards low diameter
ratios $v$ as seen in Fig. \ref{fig_sinecontrast}. The other parameters behave
qualitatively different. For the Mexican hat we find an optimum diameter
ratio $v\approx1.2\dots 1.4$ where the peak width
has a minimum and the contrast with respect to large lags shows a maximum.
For the French hat the contrast relative to large lags deteriorates
over a much broader range at low $v$ ratios and we observe an 
increased width of the peak there. Diameter ratios below $v\approx2.3$
clearly reduce the sensitivity of the French-hat filter in this case.
The increase of the peak width and the reduction of the contrast 
can be understood as resulting from 
the stronger pickup in the side lobes of the Fourier transformed
French-hat filter which are closer to the main peak and stronger for
lower $v$ ratios.

The corresponding figure computed using the mirror-continuation is
very similar. The mirror-continuation
always provides a somewhat better contrast. The effect of
contrast reduction at large lags by the filter truncation
is always higher for the French hat than for the Mexican
hat and decreasing with growing distance of the filled 
circle from the map boundaries.

The somewhat better contrast relative to small lags obtained in all cases
with the French-hat filter indicates that this filter is always more
suitable to detect a power-law scaling behaviour over a wide 
dynamic range below a dominant scale. For the detection of the
dominant scale the situation is less clear. In the sine wave 
and chess board maps the Mexican hat provides a better contrast 
but a larger peak width, for the circle map the Mexican hat
at its optimum diameter ratio is only slightly worse than the
French hat at its optimum ratio. Taking the general
uncertainties from the French-hat ripples at large lags, however,
it seems always preferable to use the Mexican-hat filter.

Comparing all results with respect to a clear 
indication of particular structure scales we find that the
Mexican-hat filter with a diameter ratio $v\le 1.4$ provides
the best resolution. Diameter ratios $v$ between 1.4 and 1.7
still produce a reasonably good sensitivity.
The French-hat filter has its maximum sensitivity
for diameter ratios $v$ between 2.3 and 2.5. Although it
produces $\Delta$-variance peaks with a smaller width than
the Mexican-hat filter, it produces in general ripple artifacts
in the $\Delta$-variance spectrum at large lags, lowering the
overall contrast of the peak, so that it should be deferred
relative to the Mexican-hat filter for the structure detection.
{\changed For non-periodic structures covered by regular maps with
rectangular boundaries, the edge treatment by mirror-continuation
gives a somewhat better contrast than the filter truncation
method but this difference is} relatively small.

\subsubsection{Spectral index}

To judge the value of the different filters with respect to the
retrieval of the correct slope of an fBm structure or a sub-map
from an fBm structure we consider two quantities: the dynamic
range over which the $\Delta$-variance slope can be reliably determined
and the difference between the actually measured
slope and the theoretical value.

The dynamic range of scales traceable with a filter of given shape
and diameter ratio is constrained by the maximum filter size that
can be used for the given map size. It is measured by the
maximum effective filter length for which the contributions from
those parts of the filter extending beyond a map boundary produce
no noticeable distortion of the
$\Delta$-variance spectrum. In the case of the French-hat filter we find 
that the diameter of the annulus, i.e. $l \times v$, may not exceed
the size of the map for reliable $\Delta$-variance values. For
the Mexican hat, the parameter $l\times v$ must not exceed $2/3$ of the
map size. From the relations between the effective filter length and
the filter size $l$ obtained in Sect. \ref{sect_effectivelength}, we see
that the dynamic range of effective lags available for a fit of
the $\Delta$-variance
spectrum grows with decreasing diameter ratio $v$.  Smaller $v$
values increase the total range of lags where the $\Delta$-variance can
be determined without being dominated by edge effects. 
Moreover, Sect. \ref{sect_scaledetect} shows that lower
$v$ ratios also tend to extend the dynamic range of scales below a structure
peak where the $\Delta$-variance follows a power law. Thus, low
ratios {\changed also seem to be favourable} for a reliable slope detection.

\begin{figure}
\epsfig{file=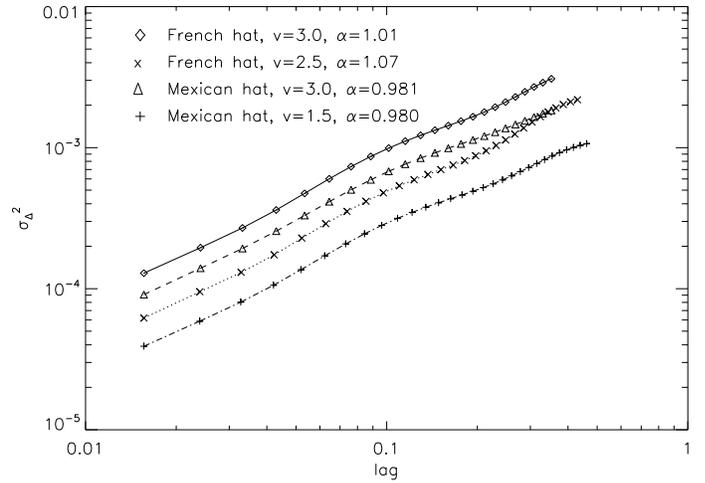, angle=90, width=\columnwidth}
\caption{$\Delta$-variance spectra computed for a sub-map from
an fBm structure with $\zeta=3$. Filter truncation is
used for the edge treatment. $\alpha$ denotes the power-law slope
fitted over the full range.}
\label{fig_cfbmdeltas}
\end{figure}

To study the agreement of the measured $\Delta$-variance slopes
with the theoretically predicted index as
a function of the filter shape we compute the 
spectra for fBm structures and sub-maps from fBm structures using the
different filters.  Fig. \ref{fig_cfbmdeltas} shows the 
spectra for a submap from an fBm with $\zeta=3.0$
computed with four different filters. {\changed First we notice
the extension of the dynamic range for lower $v$ ratios.} None of the spectra gives an
exact power law, but the French-hat filter with $v=3.0$ and both
Mexican-hat filters provide a reasonably good reproduction of the
theoretical index $\alpha=1.0$.  For low $v$ ratios, the French hat
tends to overestimate the true spectral index. The Mexican hat results in
somewhat too low exponents.

The corresponding spectra computed by the help of the mirror-continuation
method result in a spectral index which is {\changed too low by about 0.1 compared to the
theoretical value due to the flattening of the spectra at large lags
as seen in Fig. \ref{fig_edgeexample}.} The Mexican hat and the French hat provide
almost the same slopes.

When applying the analysis to the sine wave field with $k=1$, i.e. the
equivalent of an fBm with $\zeta=\infty$, the longer dynamic range
traced by the French-hat filter results in a measured spectral index
of the $\Delta$-variance which is closer to the theoretical value of 4
than that obtained for the Mexican-hat filter.

\begin{figure}
\epsfig{file=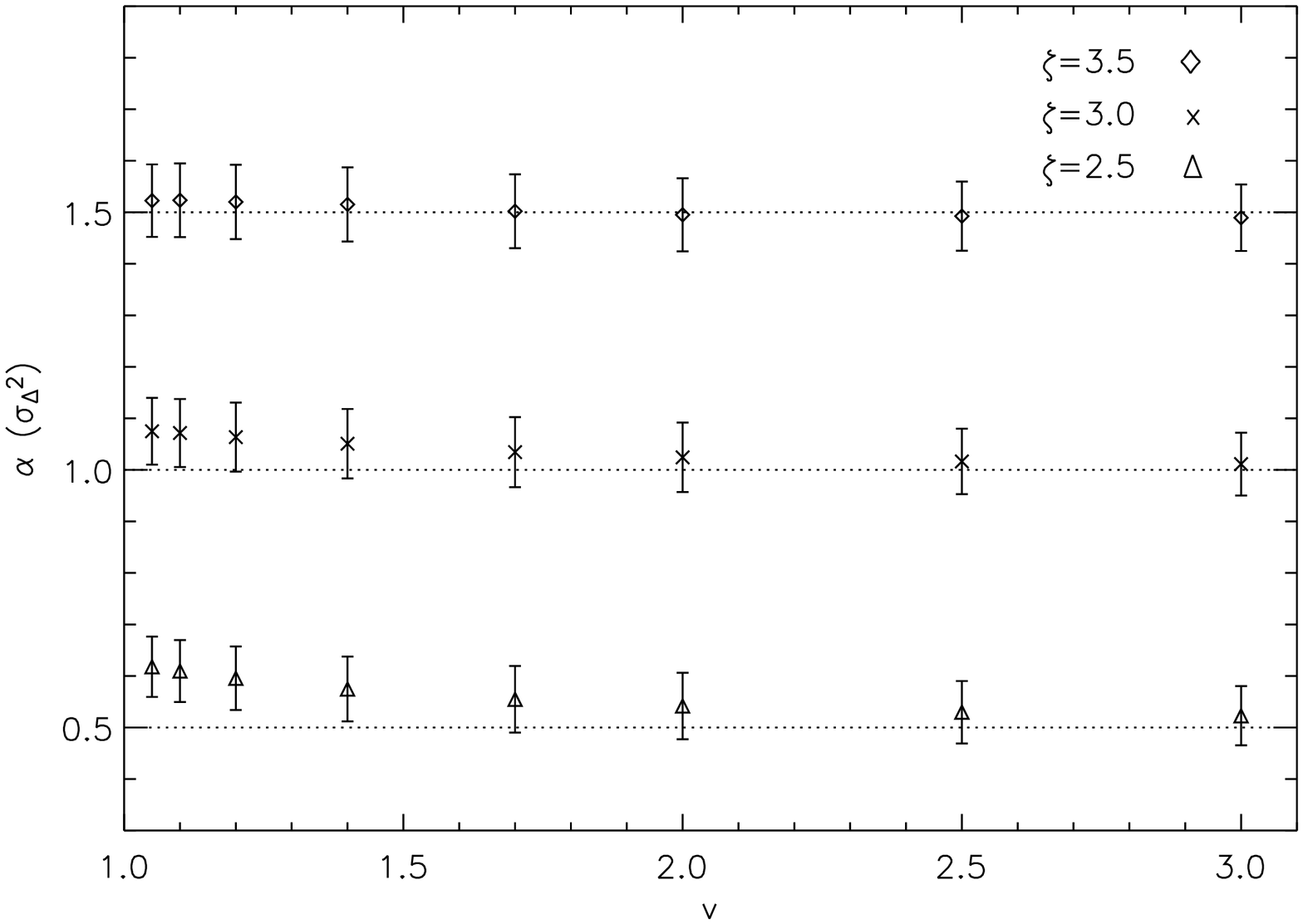, angle=90, width=\columnwidth}\vspace{0.3cm}
\epsfig{file=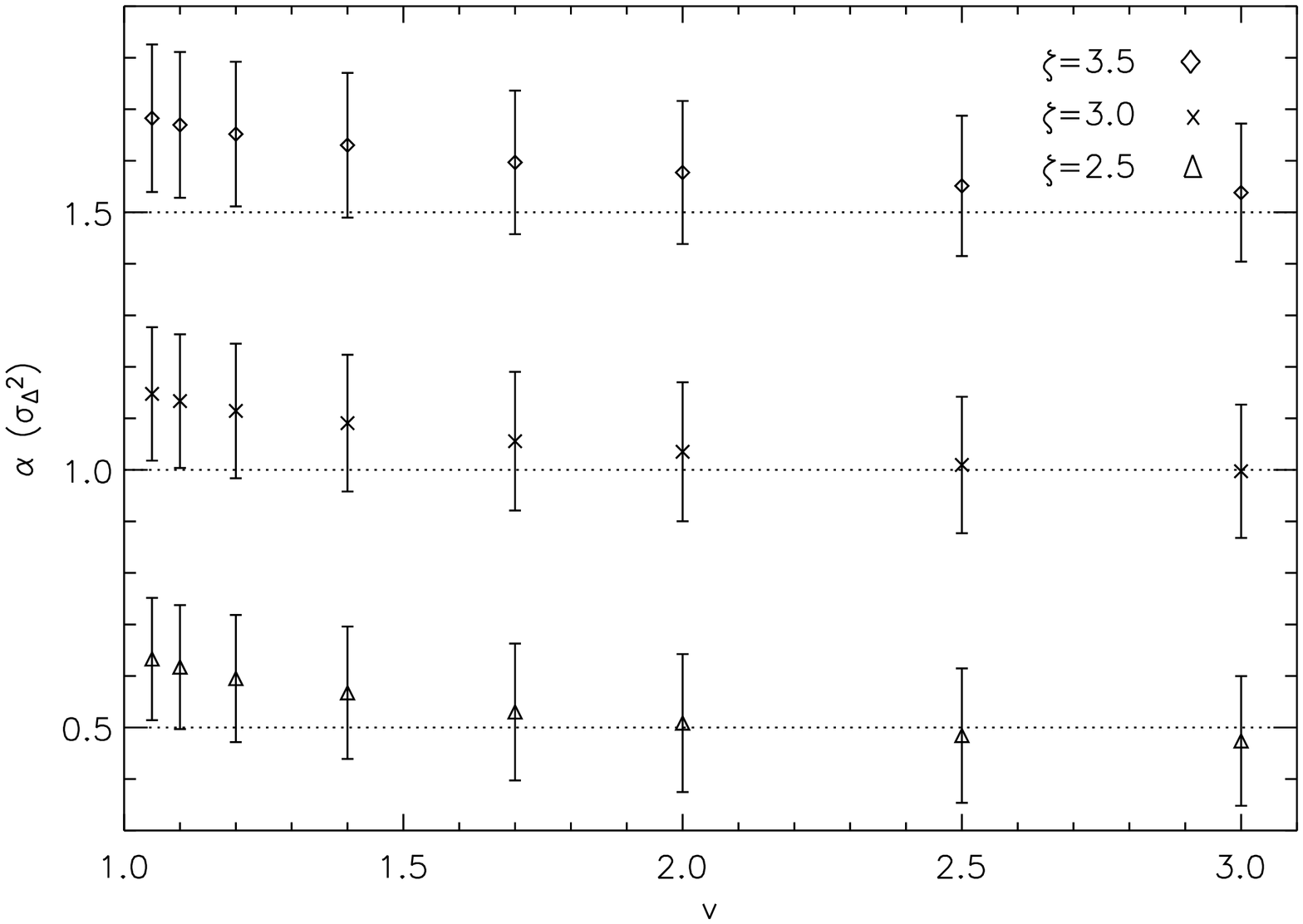, angle=90, width=\columnwidth}\vspace{0.3cm}
\caption{Distribution of fitted $\Delta$-variance exponents for 
fBm maps (upper plot) and submaps from fBm structures (lower plot)
with $\zeta=2.5$, $3.0$, and $3.5$ as a function of the filter
diameter ratio. The French-hat filter combined with the
filter truncation method is used. Dotted lines indicate the
expected $\Delta$-variance exponents corresponding to the 
fBm spectral indices. {\changed The error bars indicate the scatter
obtained for structures with the same parameters in terms of the
standard deviation.}}
\label{fig_cfbmalphas}
\end{figure}

For a systematic investigation of the accuracy in the determination of
the $\Delta$-variance slope as a function of the filter diameter ratio
$v$ we analyse sets of 30 different fBm structures and 30
submaps from fBms using the two basic filter shapes varying
their diameter ratio $v$ and the fBm spectral index $\zeta$.
Fig. \ref{fig_cfbmalphas} demonstrates the result for the French-hat
filter applied with the filter-truncation method.  The range of spectral
indices covers the typical indices in observations of
interstellar clouds \citep{Elmegreen,Falgarone04}.
The strongest deviations of the measured exponents
from the expected values occur at low diameter ratios $v$. Hence,
ratios below 1.7 should not be used.  fBms and submaps behave
differently. For the fBms the strongest deviations from the theoretical
spectral index occur at low spectral indices; for the
fBm submaps they occur at high indices. {\changed
This can be explained by selection effects. When selecting submaps
from an fBm, there is a significant scatter in the properties
of the actually selected structure. This is visible as wider error bars.
For high spectral indices we often find submaps which are
dominated by some structure extending beyond the submap boundaries.
This tends to increase the measured slope.}

Other combinations of filter shape and edge treatment show somewhat
different properties in details but the same general behaviour as
Fig. \ref{fig_cfbmalphas}.
The mirror continuation method always underestimates the
spectral index. With both filter types 
it produces at diameter ratios
$v\ge2.0$ slopes which are too low by 0.05--0.1. At lower ratios
the difference grows up to 0.2. 
The Mexican-hat filter always shows a slightly stronger deviation
from the theoretical value than the French-hat filter.
However, when systematically
correcting the slopes by a constant shift of $\Delta \alpha= 0.1$
both filter types provide reliable spectral indices for $v\ge 1.4$.
With the filter truncation method both filters result in a
good reproduction of the spectral index at diameter ratios
$v\ge 2.5$. An acceptable sensitivity is still obtained for $v \ge 1.4$
with the Mexican hat and for $v \ge 2.0$ with the
French-hat filter. In this intermediate range, the measured
average slope always deviates
by less than 0.1 from the theoretical value.

One has to keep in mind, however, that most of the deviations discussed
here fall below the size of the statistical error bars. For the
fBm submaps, which are most representative for astronomical maps,
they amount to $\Delta \alpha \approx \pm 0.15$ at a map
size of $128^2$ and to $\Delta \alpha \approx \pm 0.25$ at a map size
of $32^2$ (see Sect. \ref{sect_edgetreatment}).
Except for very low diameter ratios, $v\le 1.2$, the use of the
Mexican hat always provides a somewhat lower scattering of
the measured exponents than the French-hat filter. 
The edge treatment has practically no influence {\changed
on the size of these scatter bars}.

Comparing the results from the scale detection and the reproduction
of the spectral index we find that each problem asks 
for a different optimum filter. Whereas the scale detection 
favours the Mexican-hat filter with a diameter ratio $v\la 1.4$,
the slope reproduction is equally well satisfied by both
filter types for a diameter ratio $v\ga 2.0$.
A reasonable compromise, providing a good sensitivity to 
either issue is 
the use of a Mexican-hat filter with a diameter ratio $v\approx 1.5$.

The edge treatment by mirror continuation is somewhat favourable 
relative to the filter truncation in the slope detection, but
requires that the
average spectral index is corrected by a constant shift of $\Delta
\alpha = 0.1$.  On the other hand the filter
truncation method can be applied as well for maps with irregular
boundaries and shows a somewhat lower statistical scattering within
the studied samples. Thus, the filter truncation provides
the most reliable parameters in a single run for any data set
without the need for additional corrections.

% \section{The $\Delta$-variance calculator}
% 
% \begin{figure}
% \epsfig{file=screenshot1_bw.ps, angle=0, width=\columnwidth}
% \caption{Screen shot demonstrating the use of the IDL
% widget for the $\Delta$-variance analysis.}
% \label{screenshot}
% \end{figure}
% 
% Here, we don't use the total standard deviation of the
% filter-convolved map as the error of the $\Delta$-variance
% as proposed by \citet{Bensch} (see also erratum), because
% we found that this standard deviation
% is mainly determined by structure within the map at larger
% scales, not providing a real error of the analysis. Moreover,
% the standard deviation exceeds in many cases the total $\Delta$-variance.
% Instead we use the error of the method given by
% the lower statistically independent sampling of a given
% map when using larger filter sizes. This error term can be
% simply estimated by the Poisson error for the number of
% statistically independent filter settings at the given 
% filter size. Weighting the counting of the different filter
% settings by the reliability factors for the contributions
% within the filters then also includes the error propagation
% from the observational errors to the error in the $\Delta$-variance
% spectrum.

\section{Conclusions}

The $\Delta$-variance analysis was previously established as
a general tool to study the scaling behaviour of
interstellar cloud structure.
The main advantage of the $\Delta$-variance method compared to the 
computation of the power spectrum is its robustness with respect
to angular variations, singular distortions, gridding and finite map 
size effects.

We propose two essential improvements of the $\Delta$-variance
analysis. The first one is the use of a weighting function for each
pixel in the map. This weighting function allows us to study
data sets with a variable data reliability across the map and to 
simultaneously solve boundary problems even for maps with irregular boundaries.

Maps with a variable data reliability are eventually obtained in
most observations, either due to a local or a temporal variability
of the detector sensitivity or the atmosphere or due to 
different integration times spent for different points of a map.
The ordinary $\Delta$-variance analysis as well as the power spectrum 
fail to take the resulting effects into account.
By applying the improved $\Delta$-variance analysis to noisy data
we find that only the use of a significance function to weight
the different data points allows us to distinguish the influence
of variable noise from actual small-scale structure in the maps.
In the case of statistical uncertainties or sensitivity changes
the weighting function is best provided by the inverse RMS in the
data points. {\newchanged The weighting function allows us to 
considerably extend the dynamic range within which the intrinsic 
scaling behaviour of an observed astronomical structure 
can be measured (Appx. \ref{sect_noisetest}).}

In the treatment of map boundaries the use of a weighting function
allows us to use computational methods like the fast Fourier transform
to compute the $\Delta$-variance spectrum by extending a measured map
by points with zero significance. This is mathematically equivalent
to the truncation of the filter as proposed by \citet{Bensch}
but has the advantages of the fast computation and the definition of 
a smooth filter shape in Fourier space not influenced by gridding
effects in ordinary space. The virtual filter truncation is the only
approach to analyse maps which are only sparsely filled by significant values.
For rectangular maps a periodic continuation by mirroring can be used as
well to solve the boundary problem. Mirror-continuation will always
underestimate the spectral index by about 0.1. {\changed This could be
easily taken into account.
However, the mirror-continuation is 
less flexible than the filter truncation which works on irregular maps
as well although resulting in slightly wider statistical error bars.}

The second improvement of the $\Delta$-variance analysis
is its optimisation with respect to the shape of the wavelet used to filter
the observed maps. We have computed the effective filter length as
a function of the shape and compared the peak positions of resulting
$\Delta$-variance spectra with characteristic structure sizes in the
test data sets. We find that the peak positions always falls 10-20\,\%
below the maximum structure size. Taking this systematic offset into account
we can calibrate the spatial resolution of the $\Delta$-variance analysis
to about $\pm 5$\,\%.  Unfortunately, 
it is not possible to define a single optimum wavelet for all purposes
because different wavelets show different qualities
in the detection of the characteristic
structure scaling behaviour. The best choice for an exact measurement 
of the power spectral slope are wavelets with a high ratio between
the diameter of the annulus and the core of the filter. Here,
the French-hat and the Mexican-hat filter are equally well suited.
For the detection of pronounced size scales the Mexican-hat 
filter and low diameter ratios are preferred.
A good compromise between the
different requirements is the Mexican-hat filter with a diameter
ratio of 1.5 always providing a $\Delta$-variance spectrum with approximately
the correct slope and without missing any special spectral feature.

We provide an easy-to-use IDL widget program implementing
the {\newchanged two-dimensional} $\Delta$-variance analysis as described
here implementing the different filters and edge-treatment methods for
the analysis of arbitrary maps in FITS
format\footnote{{\tt http://www.ph1.uni-koeln.de/\~{}ossk/ftpspace/deltavar/}}.

\begin{acknowledgements}
We thank F.~Bensch for useful discussions  and J.~Ballesteros-Paredes
for carefully refereeing this paper suggesting significant
improvements. This work has been supported by the Deut\-sche
For\-schungs\-ge\-mein\-schaft through grant 494B.
It has made use of NASA's Astrophysics Data System Abstract Service.
\end{acknowledgements}

\appendix
\section{The impact of a varying data reliability}
\label{sect_noisetest}

% {\changed
{\newchanged
As a combined test of all extensions we use the results}
on the optimum filter shape with the weighting function
and return to the analysis of data with a varying reliability 
within the map.
To test how the improved $\Delta$-variance analysis
recovers the properties of an original structure from measurements
influenced by a varying noise level, we create maps where white
noise with a spatial pattern of different noise amplitudes was added.
We use combinations
of the different spatial structures discussed in Sect. \ref{sect_testdata}
for the structure to be measured and for the spatial distribution
of a noise level superimposed to the data.

\begin{figure}
\epsfig{file=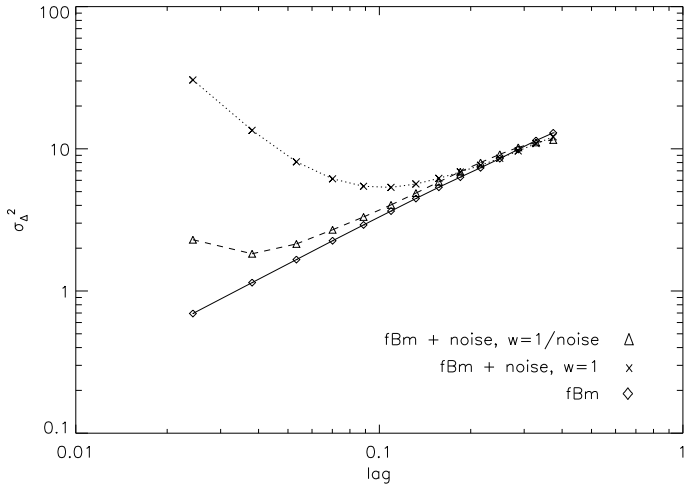, angle=90, width=\columnwidth}
\caption{$\Delta$-variance spectra computed from an fBm
structure with a spectral index $\zeta=3.1$ and a superimposed 
circular pattern of white noise. The solid line shows the
$\Delta$-variance spectrum of the original fBm, the dotted line the spectrum
obtained from the direct analysis of the noisy map, and the dashed
line the spectrum computed with a reliability weighting given
by the inverse noise RMS. }
\label{fig_circlenoise}
\end{figure}

Fig. \ref{fig_circlenoise} shows one example of a resulting $\Delta$-variance
spectrum for an fBm structure with a spectral index $\zeta=3.1$
where a noise pattern given by the filled circle described in Sect.
\ref{sect_nonperdata} and $d=2/3$ was added.  The average signal-to-noise ratio, 
defined as the ratio between the maximum in the fBm structure and the noise RMS,
is 1 and the variation
between the noise levels inside and outside of the circle is a factor 9. 
This example may represent the
situation of an observed map where the inner part is covered by
many integrations, so that it shows a high signal-to-noise ratio,
whereas the outer part is observed with few integrations leading
to a higher noise level.

The solid line represents the $\Delta$-variance spectrum of the
original fBm structure. The dotted line is the spectrum that is
obtained by the direct analysis of the noisy map without any
reliability weighting. Because there is no correlation in the
noise between neighbouring data points, the added noise contributes
to the $\Delta$-variance spectrum only on small scales with a
decay proportional to $l^{-2}$ towards larger lags. Due to the 
relatively high average noise level, the $\Delta$-variance
spectrum is dominated by the noise contribution up to lags of
about 0.2. The original fBm spectrum is only matched within a
very narrow scale range at the largest lags.

The dashed line shows the improvement that is obtained by
using the knowledge on the noise level in terms of a weighting
function $w(\vec{r})$ inversely proportional to the local
noise RMS. Due to the relative suppression of 
contributions from the outer noisy parts of the map 
the original fBm spectrum is recovered
over a much broader range of scales. We find, however, a small
distortion at the data point for the largest lag. This can
be interpreted as the effect of a slight ``cross talk''
from the weighting function to the measured structure. The
obvious strong improvement of the recovery of the original
$\Delta$-variance spectrum from the noisy data is thus
achieved at the cost of a slightly
reduced data reliability on the characteristic scales of the
weighting function.

To study this effect more systematically we perform
a number of parameter studies combining the different structures
with varying noise patterns, varying noise levels and
varying noise dynamic ranges. In the resulting 
$\Delta$-variance spectra we computed the scale range over
which the spectrum agrees with the spectrum of the original 
structure within 10\,\%.  The length of this
range, which allows a reliable derivation of the true scaling
behaviour, is a measure for the quality of the structure recovery.
To test the influence of selection effects
we repeat each computation for a number of different
initialisers for fBm structures and for the noise fields, so that
we arrive at 30--80 computations for each parameter set providing a
statistically significant sample.

\begin{figure}
\epsfig{file=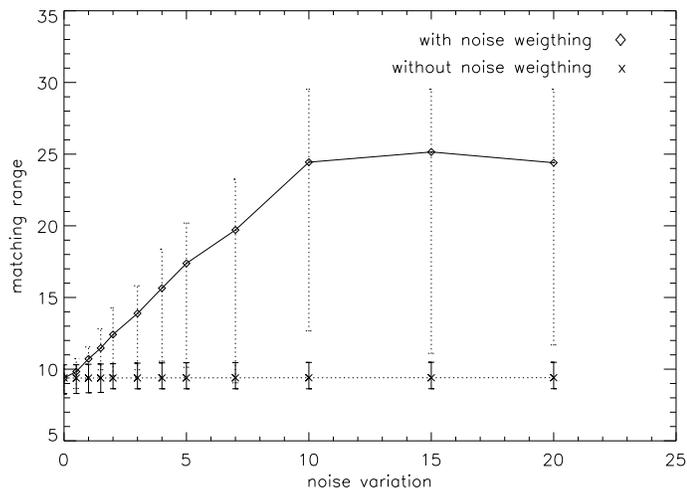, angle=90, width=\columnwidth}
\caption{Scale range (ratio between maximum and
minimum lag) within which the derived noisy spectrum agrees
within 10\,\% with the original spectrum when the $\Delta$-variance
spectrum is computed either with or without a weighting function
given by the inverse noise RMS. The
original structure is given by an fBm structure with $\zeta=3.0$
and the noise pattern by a chess board structure 
with four fields and a signal-to-noise ratio of 5.0.
On the abscissa we have changed the amplitude of the noise level
variation between the different fields of the chess board.
The error bars show the absolute minimum and maximum of the
matching range size found in a sample of 30 different fBms.
}
\label{fig_chessnoisescan}
\end{figure}

The result of such a parameter scan is demonstrated in Fig.
\ref{fig_chessnoisescan}. In this example, the
original structure is an fBm structure with $\zeta=3.0$
and the superimposed noise amplitude follows a chess board
structure (see  Sect.  \ref{sect_pertestdata}) with four fields,
i.e. a field length of half the map size. By changing the
noise amplitude ratio between the different fields while
preserving the average noise amplitude this plot is suited
to study the actual effect of the noise variation and the
corresponding correction by a weighting function in the
$\Delta$-variance analysis.

In the analysis without
weighting function, the dynamic scale range within which the
true spectrum can be fitted always covers about a factor 9,
independent of the variation in the noise level between the
four fields. The noise correction by the weighting function
can extend this range up to an average factor of 25 in the case
of high variation levels.
The error bars, indicating the minimum
and maximum ranges detected in the sample, show, however,
that there is a considerable spread in the range length over
which the fit is reliable. The matching range is always increased
compared to the $\Delta$-variance analysis without weighting function,
but the actual magnitude of this increase can considerably vary
\footnote{The upper limit at about 30 corresponds to the whole spectrum.
Thus a further extension is not possible here.}.
The example represents, however, a kind of worst case scenario,
because the noise variation in this pattern
virtually cuts out two large pieces from the
map which may contain main elements of the original structure.
This is not expected for real observations where the astronomer
would hardly select a field avoiding the main object of interest.
For a chess board noise with half the cell size,
the error bars for the distribution of fitting ranges are already
reduced by almost a factor two.
The example is nevertheless instructive because it shows all
the effects that we encounter in
the parameter study with different intensity and noise
structures. We always find the increase of the fitting range
but also the wider spread of the ranges within the
sample studied. When using fBms for the noise distribution
with spectral indices $\zeta$ around 4 or higher,
we find as well a wide spread of the factors by which the length of the 
fitting range is increased in the new $\Delta$-variance analysis.
For all lower indices the error bars shrink in
the same way as for the chess board with smaller cell size.
In general, the extension of the matching range by the use of a 
weighting function is most significant for strongly fragmented
noise maps and noise maps well adapted to the dominant parts of 
the actual structure. The latter case is
usually given in astronomical observations.

We conclude that the introduction of a weighting function
given by the inverse noise of the data into the $\Delta$-variance
analysis always extends the spatial range over which the original
scaling behaviour can be recovered. The amount of this improvement
depends on the strength of the noise variation across the map
and the coverage of the observed structure by regions of low noise.
The range within which the measured $\Delta$-variance
spectrum agrees with the original spectrum can be extended by 
more than a factor three for high levels of noise variations and
a good coverage of the main features of a structure by low-noise observations.
Future studies should show whether 
the information from the $\Delta$-variance spectrum of the
map of weights can be used to further improve the outcome.
% }

\end{document}